# A Lower Limit of Atmospheric Pressure on Early Mars Inferred from Nitrogen and Argon Isotopic Compositions

Hiroyuki Kurokawa[a], Kosuke Kurosawa[b], Tomohiro Usui[a,c]

[a]Earth-Life Science Institute, Tokyo Institute of Technology, 2-12-1 Ookayama, Meguro-ku, Tokyo 152-8550, Japan

[b] Planetary Exploration Research Center, Chiba Institute of Technology, 2-17-1, Tsudanuma, Narashino, Chiba 275-0016, Japan

[c]Department of Earth and Planetary Sciences, Tokyo Institute of Technology, 2-12-1 Ookayama, Meguro-ku, Tokyo 152-8551, Japan

## Abstract

We examine the history of the loss and replenishment of the Martian atmosphere using elemental and isotopic compositions of nitrogen and noble gases. The evolution of the atmosphere is calculated by taking into consideration various processes: impact erosion and replenishment by asteroids and comets, atmospheric escape induced by solar radiation and wind, volcanic degassing, and gas deposition by interplanetary dust particles. Our model reproduces the elemental and isotopic compositions of N and noble gases (except for Xe) in the Martian atmosphere, as inferred from exploration missions and analyses of Martian meteorites. Other processes such as ionization-induced fractionation, which are not included in our model, are likely to make a large contribution in producing the current Xe isotope composition. Since intense impacts during the heavy bombardment period greatly affect the atmospheric mass, the atmospheric pressure evolves stochastically. Whereas a dense atmosphere preserves primitive isotopic compositions, a thin atmosphere on early Mars is severely influenced by stochastic impact events and following escape-induced fractionation. The onset of fractionation following the decrease in atmospheric pressure is explained by shorter timescales of isotopic fractionation under a lower atmospheric pressure. The comparison of our numerical results with the less fractionated N ($^{15}N/^{14}N$) and Ar ($^{38}Ar/^{36}Ar$) isotope compositions of the ancient atmosphere recorded in the Martian meteorite Allan Hills 84001 provides a lower limit of the atmospheric pressure in 4 Ga to preserve the primitive isotopic compositions. We conclude that the atmospheric pressure was higher than approximately 0.5 bar at 4 Ga.

*Keywords:* Mars, atmosphere: Atmospheres, evolution: Atmospheres, composition

# 1. Introduction

Whereas present-day Mars has a thin atmosphere (6 mbar on average), ancient Mars probably had a denser atmosphere. Geomorphological evidence such as valley networks and deltas on Mars requires repeated episodes of liquid water runoff in the Noachian period (e.g., Hynek et al. 2010). A dense atmosphere with a climate warm enough to sustain liquid water on the Martian surface has been proposed to explain the fluvial terrains (e.g., Pollack et al. 1987; Forget and Pierrehumbert 1997; Ramirez et al. 2014).

The idea of a permanent warm climate on early Mars was questioned; climate models showed that a dense $CO_2$ atmosphere on Mars cannot maintain a mean surface temperature higher than the freezing point of water (e.g., Kasting 1991; Forget et al. 2013), and geochemical evidence suggested that water fluvial activity was mostly limited to the subsurface (Ehlmann et al. 2011). The ancient Mars might have been cold, but global atmospheric circulation under a dense atmosphere is needed to transport water to highlands by atmospheric circulation, and consequently to create fluvial terrains by episodic melting events of the ice on the highlands, even in the cold early Mars scenario (Wordsworth et al. 2013; 2015; Cassanelli and Head 2015).

The Martian atmosphere would have been lost and replenished by several processes (Figure 1). Exploration missions and analyses of Martian meteorites have revealed that almost all the volatile elements in the Martian atmosphere and surficial water are enriched in heavier isotopes, which suggests that lighter isotopes were preferentially removed by atmospheric escape induced by solar radiation and wind (Jakosky and Phillips 2001). In addition, impacts of asteroids and comets erode and replenish the atmosphere (Melosh and Vickery 1989; Zahnle 1993; de Niem et al. 2012). Volcanic degassing (Craddock and Greeley 2009) and gas deposition by interplanetary dust particles (IDPs) (Flynn 1997) would have contributed to the replenishment of the atmosphere. Although the escape/replenishment processes are not completely understood, the geological evidence mentioned above suggests a net decrease in the atmospheric pressure throughout the Martian history.

Elemental and isotopic compositions of the atmosphere have been used to trace the history of the Martian atmosphere in literature (Jakosky 1991; Jakosky et al. 1994; Pepin 1991; 1994; Jakosky and Jones 1997; Zahnle 1993; Slipski and Jakosky 2016). Out of these, Jakosky et al. (1994) and Pepin (1994) are the most comprehensive studies that examined the evolution of the Martian atmosphere, taking atmospheric escape induced by solar radiation and wind, impact erosion, and volcanic degassing into account. They concluded that the fractionated isotopic compositions of N, Ne, and Ar in the Martian atmosphere are explained by interplay between the escape and replenishment by volcanic degassing. Contrary to the unfractionated Kr isotopes, the fractionated Xe was suggested to be the remnants of a hydrogen-rich primordial atmosphere lost by hydrodynamic escape at the earliest stage of the Martian evolution (Pepin 1994).

Whereas these comprehensive studies are suggestive to constrain the evolution of the Martian atmosphere, recent isotopic measurements of the Martian atmosphere by *Curiosity* (Atreya et al. 2013; Wong et al. 2013) and analyses of the Martian meteorites (Mathew and Marti 2001) provided new data that can be compared with model calculations. In particular, $^{15}N/^{14}N$ and $^{38}Ar/^{36}Ar$ ratios of the early atmosphere recorded in Martian meteorite Allan Hills (ALH) 84001 (Mathew and Marti 2001) are used to constrain the atmospheric pressure on early Mars at 4.1 Ga (the crystallization age of ALH 84001, Bouvier et al. 2009; Lapen et al. 2010) in this study.

We constrain the atmospheric pressure on early Mars by combining a numerical model and isotopic data recently obtained from exploration missions and from analyses of Martian meteorites. Our numerical model is explained in Section 2. In Section 3, we calibrate input parameters to reproduce the elemental and isotopic composition of the present-day atmosphere and show a constraint on the atmospheric pressure on early Mars by comparing the numerical results with the isotopic data. The evolution of the Martian surface environment is discussed in Section 4. Finally, we conclude that the atmospheric pressure was higher than ~0.5 bar at 4 Ga in Section 5.

# 2. Model

We constructed a one-box atmosphere-hydrosphere/cryosphere model containing multiple species: $CO_2$, $N_2$, $H_2O$, and noble gases (Figure 1). We calculated the evolution of the elemental and isotopic compositions of the Martian atmosphere, taking into consideration the escape and replenishment processes: impact erosion and replenishment by asteroids and comets, atmospheric escape induced by solar radiation and wind, volcanic degassing, and gas deposition by IDPs.

## 2.1. Climate model and phase equilibrium

Phase equilibrium of $CO_2$ and $H_2O$ between the atmospheric and solid/liquid reservoirs is treated in our model. The Martian atmosphere exchanges $CO_2$ with other $CO_2$ reservoirs on the surface: ice caps and regolith (Gierasch and Toon 1973). Depending on the atmospheric pressure, the atmosphere can collapse to form large polar $CO_2$ ice caps because of the instability (Haberle et al. 1994; Kurahashi-Nakamura and Tajika 2006). Recent three-dimensional global-circulation-model simulations (Forget et al. 2013) showed that $CO_2$ ice caps are formed when the surface pressure is higher than 3 bar. There is another threshold at lower atmospheric pressure; if the atmosphere is thinner than this threshold, the atmosphere collapses to form large $CO_2$ ice caps. This threshold value depends on the obliquity of Mars (Nakamura and Tajika 2003; Forget et al. 2013). The Martian obliquity is chaotic in the timescale of Myrs, at least for the current condition (Laskar et al. 2004)[1]. A threshold of the atmospheric collapse, $P_{collapse}$ = 0.5 bar, was assumed in most of our model by following the maximum value obtained by Forget et al. 2013. Since $CO_2$ is the dominant species in the Martian atmosphere, the abundances of N and noble gases in the atmosphere would increase significantly after the collapse, which leads to the following escape-induced fractionation (Jakosky et al. 1994; Pepin 1994, Section 3). We will also show the case without the atmospheric collapse in subsection 3.2.

---

[1] The obliquity could be stable in the Noachian period depending on the configuration of giant planets in the Nice model (Ramon and Walsh 2011)

Water vapor is another condensation component in the Martian atmosphere. The atmospheric $H_2O$ pressure $p_{H_2O}$ is given by,

$$p_{H_2O} = \exp\left(9.550426 - \frac{5723.265}{T} + 3.53068 \ln T - 0.00728332\, T\right) \quad \text{[Pa]} \qquad (1)$$

where $T$ is the temperature (Murphy and Koop 2005). We assume $T = 273.16$ K ($H_2O$ triple point) considering a possible warmer climate and $T = 210$ K (mean surface temperature on present-day Mars, e.g., Barlow 2008) before and after the collapse. Our model is simplified by employing a constant value for $p_{H_2O}$; in reality, the amount of water vapor in the atmosphere varies as a function of the obliquity (Mellon and Jakosky 1995; Mellon and Phillips 2001). We have done sensitivity analyses and found that the effect of $p_{H_2O}$ on the atmospheric evolution is minor, as long as the abundance of water vapor is lower than 1%. The atmospheric $CO_2$ pressure after the collapse is controlled by surface $CO_2$ reservoirs and is assumed to be 6 mbar (mean surface pressure on present-day Mars, e.g., Jakosky and Phillips 2001). We note that carbonate formation is not included in our model, but we discuss its influence in Section 4.

## 2.2. Impacts of asteroids and comets

The impacts of asteroids and comets both remove and replenish atmospheric volatiles (Pham et al. 2009; 2012; de Niem et al. 2012). A Monte Carlo method (Kurosawa 2015) was used to treat the stochastic nature of impacts in our model by assuming the total mass of the impactors, size distribution, velocity distribution, and impact flux. The *Mersenne Twister* algorithm (Matsumoto and Nishimura 1998) was used in the Monte Carlo calculation.

Atmospheric loss and volatile delivery of each impact are calculated by adapting empirical formulas obtained by the recent numerical simulations (Shuvalov 2009). Since the total volatile abundance in an asteroid depends on the asteroidal types (Wasson and Kallemeyn 1988), it is treated as a parameter $X_{gas}$ (defined by the ratio of the degassed $CO_2$ mass to the total mass of the asteroid). We used $X_{gas} = 10\%$ (carbonaceous chondrite-like) and 1% (enstatite chondrite-like) based on the carbon content of meteorites (Wasson and Kallemeyn 1988). The masses of other volatiles are calculated by assuming that the relative abundances to C are chondritic (Figure 2). Some scenarios predict cometary impacts during the late heavy bombardment (LHB) period (e.g., Gomes et al. 2005). We treated the ratio of comets to asteroids after 4.1 Ga (see the next paragraph) as a parameter $f_{comet}$ (the number ratio of comets to asteroids in the impactors). We calculated the cases of $f_{comet} = 0\%, 0.1\%$, and 1%. It will be shown that $f_{comet}$ should be less than 1% to reproduce the abundances of noble gases on the present-day Mars (Section 3).

We assumed the total mass of impactors to be $2 \times 10^{21}$ kg, which was estimated by assuming that the Martian mantle abundances of the highly siderophile elements (HSEs) are similar to those in the Earth's mantle (Bottke et al. 2010). The size distribution of the impactors was assumed to follow the power laws ($\mathrm{d}N/\mathrm{d}D \propto D^{-q}$, where $N$ is the cumulative number of the impactors and $D$ is the diameter of the impactor). We assumed $q = 2$ for large impactors (D > 300 km), which is estimated to reproduce the mantle abundances of HSEs in Earth and the Moon (Bottke et al. 2010), and $q = 3.5$ for small impactors (D < 300 km), which is the value expected from the collisional cascade (Tanaka et al. 1996) as shown in Figure 3a. The velocity distribution of the impactors is adapted from that of the asteroid belt (rescaled to the impact on Mars from Ito and Malhotra 2006, Figure 3b). We first calculated the total number of impacts from the assumed total mass and size distribution. Impact flux as a function of the time was then obtained from the calculated total number of impacts and a crater chronology model (Morbidelli et al. 2012). We note that the timing of the LHB was assumed at ~ 4.1 Ga in the chronology model. Impacts are mostly concentrated prior to 4 Ga (so-called late accretion, which includes the LHB).

We adopted empirical formulas for the loss of the atmospheric mass $\Delta m_{\mathrm{atm}}$ and the projectile mass $\Delta m_{\mathrm{pr}}$ shown by Shuvalov (2009). The net loss (gain) of an atmospheric species "i" is written as, $\Delta m_i = X_{\mathrm{i,atm}} \Delta m_{\mathrm{atm}} + X_{\mathrm{i,pr}}(m_{\mathrm{pr}} - \Delta m_{\mathrm{pr}})$, where $m_{\mathrm{pr}}$ and $m_{\mathrm{atm}}$ are the masses of the projectile (impactor) and atmosphere, and $X_{\mathrm{i,pr}}$ and $X_{\mathrm{i,atm}}$ are the mass abundances in the projectile and atmosphere. The loss and gain are described using a dimensionless variable $\xi$,

$$\xi = \frac{D^3 \rho_t \rho_{\mathrm{pr}}(V^2 - u_{\mathrm{esc}}^2)}{H^3 \rho_0 u_{\mathrm{esc}}^2 (\rho_t + \rho_{\mathrm{pr}})} \quad , \qquad (2)$$

where $\rho_{pr}$ and $\rho_t$ are the densities of projectile (impactor) and target (Martian surface), $\rho_0$ is the atmospheric density at the surface, $H$ is the atmospheric scale height, and $V$ is the impact velocity. The normalized mass lost from the atmosphere,

$$X_{\mathrm{a}} \equiv \frac{\Delta m_{\mathrm{atm}}}{m_{\mathrm{atm}}} \frac{u_{\mathrm{esc}}^2}{V^2 - u_{\mathrm{esc}}^2}, \qquad (3)$$

is given by,

$$\log X_{\mathrm{a}} = -6.375 + 5.239 \log \xi - 2.121 \left(\log \xi\right)^2$$
$$+ 0.397 \left(\log \xi\right)^3 - 0.037 \left(\log \xi\right)^4 + 0.0013 \left(\log \xi\right)^5. \qquad (4)$$

The normalized projectile mass lost, $X_{\mathrm{pr}} \equiv \Delta m_{\mathrm{pr}}/m_{\mathrm{pr}}$, is given by,

$$X_{\mathrm{pr}} = \min(0.035 \frac{\rho_t}{\rho_{\mathrm{pr}}} \frac{V}{u_{\mathrm{esc}}} (\log \xi - 1), 0.07 \frac{\rho_t}{\rho_{\mathrm{pr}}} \frac{V}{u_{\mathrm{esc}}}, 1). \qquad (5)$$

The volatile abundances in asteroids and comets are shown in Figure 2. The volatile abundance in chondrites depends on the chondrite types. We assumed the relative abundance (the ratio of N and noble gases to C) of CI chondrites. Then the absolute abundances of volatiles in asteroids are determined by a parameter $X_{gas}$. The noble gas abundance in comets is taken from Dauphas (2003) (the 50 K model in their Table 1). The C and N abundances are then estimated by assuming the solar C/Ar and C/Ar ratios by following the assumption used in Marty and Meibom (2007). We will show that noble gases in the Martian atmosphere are mostly supplied from comets and IDPs (Section 3).

*2.3. Sputtering*

Oxygen atoms in the upper atmosphere are ionized by extreme ultraviolet (EUV) photons, electron impacts, or charge exchange. The $O^+$ ions are picked up by the magnetic field of the solar wind. The picked-up ions re-impact the particles in the upper atmosphere because their gyration radii are comparable with the size of Mars. These ions sputter atmospheric molecules and atoms away (e.g., Luhmann et al. 1992).

Sputtering followed the model used in Jakosky et al. (1994) and Pepin (1994) with some modifications. The $CO_2$ sputtering rate $F_{CO_2, sp}{}^{tot}$ calculated by the $CO_2$ + O exobase model (Luhmann et al. 1992) for 1, 3, and 6 times present-day EUV flux was adapted by interpolating and extrapolating as a function of time. The sputtering rate of each species was calculated from $F_{CO_2, sp}{}^{tot}$ as,

$$F_{i, sp} = F_{CO_2, sp}{}^{tot} \frac{Y_i}{Y_{CO_2}} \frac{N_i}{N_{CO_2}} R_{diff, i/CO_2} 1/\alpha, \qquad (6)$$

where $F_{i, sp}$ is the sputtering rate, $Y_i$ is the yields, $N_i$ is the number density at the homopause, $R_{diff, i/CO_2}$ is the fractionation factor by diffusive separation between the homopause and exobase, and the dimensionless factor $\alpha$ is defined by $\alpha \equiv \sum_i \frac{N_i}{N_{CO_2}} R_{diff, i/CO_2}$, which denotes the summation of the mixing ratios at the exobase. The fractionation by diffusive separation is given by,

$$R_{diff, i/j} = \exp(-\Delta m_{i,j} g \Delta z / k_B T), \qquad (7)$$

where $\Delta m_{i,j}$ is the mass difference between two species (j is $CO_2$ for $R_{diff, i/CO_2}$), $g$ is the gravitational acceleration, $\Delta z$ is the distance between homopause and exobase, and $k_B$ is the Boltzmann constant. We adopted the yields calculated by Monte Carlo particle simulations for N, Ne, and Ar (Table 1, Jakosky et al. 1994). Because the yields of Kr and Xe were not provided in Jakosky et al. (1994), these values were estimated from an analytical formula shown by Johnson (1992) (Table 1). We note that the $CO_2$ sputtering rate in our multiple component model $F_{CO_2, sp}$ is not identical to that obtained by the $CO_2$ + O exobase model ($F_{CO_2, sp}{}^{tot}$) because of the factor $1/\alpha$ which denotes the dilution effect by other species.

The time dependence of the solar EUV flux and the model of the upper atmosphere were updated from the model of Jakosky et al. (1994) and Pepin (1994). The model of the EUV flux obtained from observations of solar proxies in different ages (Ribas et al. 2005) was used. The saturation phase was assumed for the first 0.1 billion years (Jackson et al. 2012) of the solar evolution (the first 0.04 billion years in our calculation). We adopted the upper atmospheric structures under different EUV levels for $\Delta z/T$ (Terada et al. 2009; Zhang et al. 1993; Jakosky et al. 1994, Table 2) and interpolated through time.

A global magnetic field produced by ancient Martian dynamo might prevent the atmosphere from being sputtered (Jakosky and Phillips, 2001). The magnetic signatures of impact basins on Mars suggest that the dynamo ceased at ~ 4.1 Ga (Lillis et al. 2008). We assumed two end members. The sputtering is assumed to operate from 4.1 Ga in most of our models. The case where the sputtering operated from the beginning of our calculation will be discussed in Section 4.

*2.4. Photochemical escape*

Nitrogen escapes from the upper atmosphere by photochemical processes including photodissociation, photodissociative recombination, electron impact dissociative ionization, dissociative recombination, and ion-molecule reactions (Fox and Dalgarno 1983; Fox 1993). Photochemical escape of N followed the model used in Jakosky et al. (1994) and Pepin (1994) with some modifications: the time dependence of the solar EUV flux (subsection 2.3) and the fractionation factor. The escape rate depends on the $N_2/CO_2$ mixing ratio and solar EUV flux. The dependence on the $N_2/CO_2$ mixing ratio was taken from Fox and Dalgarno (1983). The dependence on the EUV flux $F_{EUV}$ was assumed by following the discussion of Jakosky et al. (1994): the escape rates by molecule-photon and molecule-electron reactions (photodissociation, photodissociative recombination, electron impact dissociative ionization, and dissociative recombination) are proportional to $F_{EUV}$ (Fox and Dargarno, 1983; Fox 1993) and the escape rates by ion-molecule reactions are proportional to $F_{EUV}^{1.5}$.

Carbon is also removed from the upper atmosphere by photochemical processes including photodissociation of CO and dissociative recombination of $CO^+$ and $CO_2^+$ (Groeller et al. 2014; Hu et al. 2015), which are included in our model as a mechanism of $CO_2$ loss. The model used in Hu et al. (2015) was adopted in our model. The photochemical escape rate $F_{c.ph}$ is given by,

$$F_{c.ph}=7.9 \times 10^{23} F_{Lyc}^{\beta} \quad [s^{-1}], \qquad (8)$$

where $F_{Lyc}$ is the solar Lyman continuum flux in units of the current solar Lyman continuum flux ( $F_{Lyc} \propto t^{0.86}$, where $t$ is the time) and $\beta$ is the power-law index. The index $\beta$ was treated as a parameter in Hu et al. (2015). We show the cases of $\beta = 1$ in the following sections, but the results do not depend on the assumption because the influence of the photochemical escape on the evolution of the atmospheric pressure is minor.

## 2.5. Volcanic degassing

A volatile supply by volcanic degassing is likely to be one of the major processes of the atmospheric replenishment through Martian history. The dependence of the degassing rate on time was taken from Craddock and Greeley (2009), which is based on geologic records of volcanism on Mars. Craddock and Greeley (2009) showed the minimum estimate of the gases released into the atmosphere by using the volatile abundance of Earth's basaltic eruptions. The degassing composition is shown in Figure 2. The model of Craddock and Greeley (2009) was adopted for $CO_2$, $H_2O$, and $N_2$ (Figure 4). We estimated the abundance of the noble gases in the volcanic gas by assuming their ratios to (C+N) to be identical to those of the Earth's depleted mantle (Marty 2012). The cumulative amount of degassed $CO_2$ during the Martian history corresponds to 0.4 bar in the Craddock and Greeley (2009) model. We introduce the volcanic factor $C_{vol}$, which is a multiplicative factor that accounts for a different volcanic rate, and is the same approach employed by Slipski and Jakosky (2016). We calculated the cases of $C_{vol} = 1, 3, 5$. We will show that the volcanic degassing is a major source of N in the Martian atmosphere, whereas the noble gases mostly originated from comets and IDPs (Section 3).

## 2.6. Gas deposition by interplanetary dust particles

The origins of IDPs are thought to be comets and asteroids. Because IDPs are abundant in noble gases, it has been proposed that IDPs are a non-negligible source of the atmospheric Ne in the Martian atmosphere (Flynn 1997). We adopted a gas deposition model proposed by Flynn (1997) for the noble gases (Table 3 and Figure 2). The gas deposition rate is assumed to be constant through time in our model. Because Ne abundance in IDPs varies in samples (Pepin et al. 2000) and higher abundance than the IDP samples was reported from cometary samples obtained by *Stardust* mission (Marty et al. 2008), we also tested cases where the rate of Ne deposition is an order of magnitude higher than the Flynn (1997) model. We refer to the Ne abundance in IDPs normalized by the Flynn (1997) model as $C_{Ne,IDP}$. We calculated the cases of $C_{Ne,IDP} = 1$ and 10. We will show that the higher Ne abundance is suitable to explain the Ne abundance in the Martian atmosphere (Section 3). We note that in this study it is assumed that the gas deposition from IDPs occurs immediately after injection into the atmosphere.

## 2.7. Isotopic fractionation

Isotopic compositions are changed by the escape-induced fractionation, continuous supplies from volcanic degassing and IDPs, and stochastic impact replenishments. The change rate caused by the escape is written as,

$$\left(\frac{dI_{m,n}}{dt}\right)_{escape} = I_{m,n}(R_{m/n}-1)\frac{\sum_k N_k}{\sum_k R_{m,k} N_k}\frac{(dN/dt)_{escape}}{N}, \qquad (9)$$

where m and n denote two isotopes, $I_{m,n}$ is the isotope ratio, $R_{m/n}$ is the fractionation factor, $N$ is the number density of the element considered, and $N_k$ is the number density of the isotope $k$. The change rate caused by the continuous supplies is given by,

$$\left(\frac{\mathrm{d}I_{\mathrm{m,n}}}{\mathrm{d}t}\right)_{\mathrm{supply}} = (I_{\mathrm{m,n}}{}^0 - I_{\mathrm{m,n}}) \frac{\sum_k I_{\mathrm{m,k}}}{\sum_k I_{\mathrm{m,k}}{}^0} \frac{(\mathrm{d}N/\mathrm{d}t)_{\mathrm{supply}}}{N}, \qquad (10)$$

where superscript 0 denotes the value of the source. Finally, the isotope ratio after impact replenishment is given by,

$$I_{\mathrm{m,n}}{}' = \frac{I_{\mathrm{m,n}}(\sum_k I_{\mathrm{m,k}}{}^0)N + I_{\mathrm{m,n}}{}^0(\sum_k I_{\mathrm{m,k}}) \Delta N}{\sum_k I_{\mathrm{m,k}}{}^0 N + \sum_k I_{\mathrm{m,k}} \Delta N}, \qquad (11)$$

where $I_{\mathrm{m,n}}{}'$ is the isotope ratio $I_{\mathrm{m,n}}$ after the impact and $\Delta N$ is the change in the number density of the element considered. We note that the atmospheric erosion caused by the impacts does not affect the isotope compositions.

Because the sputtering is an energetic process, the fractionation is caused only by diffusive separation between the homopause and exobase. The fractionation factor $R_{\mathrm{diff,m/n}}$ is given by Equation 7. The photochemical escape has an additional fractionation effect and the fractionation factor for N $R_{\mathrm{ph,14N/15N}}$ is 1.3-1.4 $\times$ $R_{\mathrm{diff,14N/15N}}$ (Wallis 1989; Fox and Hác 1997; Manning et al. 2008). We assumed $R_{\mathrm{ph,14N/15N}} = 1.4 \times R_{\mathrm{diff,14N/15N}}$.

*2.8. Isotopic compositions*

We summarized isotopic compositions used in our model in Table 4. The N isotopic composition in the Martian interior was estimated from the Martian meteorites ($\delta^{15}$N = -30‰, Mathew and Marti 2001). The Ne isotopic composition of volcanic gas is assumed to be solar (Pepin 1991) from considering the presence of solar-type Ne in the Earth's mantle (Marty 2012). Other volcanic noble gases are assumed to be chondritic (Pepin 1991). Our results will show that the atmospheric noble gases are likely to have originated from comets and IDPs. Therefore, the assumption about isotopic compositions of volcanic noble gases does not affect our results.

Isotopic compositions of volatiles in asteroids are assumed to be chondritic (Pepin 1991). The N isotopic composition, which gives a lower limit of the atmospheric pressure on early Mars in Section 3, varies from ∼ -30‰ (enstatite chondrites) to ∼40 ‰ (CI chondrites). We assumed $\delta^{15}$N = -30‰, which is identical to the $\delta^{15}$N value of the Martian building blocks preserved in the Martian mantle (Mathew and Marti, 2001), to obtain a lower limit of the atmospheric pressure on early Mars by using the N isotopes (Section 3).

The N isotopic composition in comets is assumed to be $\delta^{15}$N = 1000‰, which is a typical value obtained from telescopic observations (Füri and Marty 2015). We adopted the $^{38}$Ar/$^{36}$Ar ratio recently measured by *Rosetta* mission (Balsiger et al. 2015) and assumed solar composition for other noble gases (Pepin 1991, Dauphas 2003).

Solar composition is assumed for the noble gases in IDPs (Pepin 1991). As shown in Section 3, only the atmospheric Ne possibly originated from IDPs. Therefore, the assumption on the noble gas isotopes except for Ne in IDPs does not affect our results.

We note that $^{40}$Ar has a radiogenic origin (radioactive decay of $^{40}$K) and is not included in our model. A crustal erosion model was proposed to account for the evolution of $^{40}$Ar, but the erosion process is poorly constrained (Slipski and Jakosky 2016). Incorporating the crustal erosion will impose another free parameter on us and therefore we decided not to include this isotope with a different origin.

*2.9. Initial conditions*

All of our calculations start at the end of the early hydrodynamic-escape stage at ~4.5 Ga when the primordial hydrogen-dominated atmosphere has been lost (Pepin 1994). Thus, we modeled the atmospheric history from 4.5 Ga to the present in this study. The validity of this assumption will be discussed in subsection 4.3. The initial atmospheric pressure is unknown, but it should be less than 3 bar, as a denser $CO_2$ atmosphere is unstable against the collapse (Forget et al. 2013). We assumed 1.5 bar as the initial pressure in most of our calculations. Because the stochastic behavior of impacts provides various evolution tracks from this specific initial condition (see Section 3), our conclusion does not depend on the initial pressure as long as it is several bars.

The isotopic compositions of the atmosphere in the initial state correspond to those after the hydrodynamic escape of the primordial hydrogen-rich atmosphere (Pepin 1991; 1994). We assumed volcanic isotopic compositions for N and noble gases except for Xe. Because the fractionated Xe of the Martian atmosphere has been interpreted as being a remnant of the primordial atmosphere (Pepin 1991; Pepin 1994), its present-day isotopic composition was utilized to test the hypothesis.

# 3. Results and data interpretations

First, we will calibrate our model by comparing the model outputs to the elemental and isotopic compositions of the Martian atmosphere (subsection 3.1). We will constrain input parameters ($X_{gas}$, $C_{vol}$, $C_{Ne, IDP}$, and $f_{comet}$) by examining the relation between individual inputs and outcomes rather than exploring the entire parameter space. It will be shown that the atmospheric pressure governs the fractionation regime of isotopes in the atmosphere (subsection 3.2). The calibrated model will be used to constrain the atmospheric pressure on early Mars by comparing our results to the isotopic data of the Martian atmosphere at 4.1 Ga recorded in the Martian meteorite ALH 84001 (subsection 3.3).

*3.1. Model calibration: reproducing the present-day atmosphere*

We show the evolution of atmospheric pressure in our baseline model in Figure 5. The baseline model assumed $f_{comet} = 0$, $C_{vol} = 1$ and $C_{Ne,IDP} = 1$. The results are shown for $X_{gas} = 10\%$ (Figure 5a) and 1% (Figure 5b). Since intense impact bombardment during the first ~ 1 Gyrs largely affects the atmospheric mass, the atmospheric pressure evolved stochastically. Different evolutionary tracks were obtained from a hundred Monte Carlo simulations. The results showed a dominance of accumulation when volatile-rich asteroids were assumed as impactors (Figure 5a). It is consistent with the results of de Niem et al. (2012). In contrast, in cases where relatively volatile-poor impactors were assumed, the impact erosion prevailed over the replenishment of volatiles and led to atmospheric collapse occurring at ~4 Ga (Figure 5b). The net decrease of the atmospheric pressure over time is consistent with the geological evidence (Section 1). Hereafter we assumed $X_{gas}$ = 1%. We note that this assumption—that the late accretion was dominated by volatile-poor impactors—is supported by recent isotopic analyses of the late veneer in the Martian mantle (Fischer-Gödde and Kleine 2017).

Whereas the primitive isotopic compositions were preserved for a long time under a dense atmosphere, a thin atmosphere is influenced by the escape and replenishment processes on a short timescale (subsection 3.2). Thus, the different evolutionary tracks in Figure 5 have different isotopic signatures. We compare the simulated atmospheric compositions to the elemental and isotopic data of the present-day atmosphere obtained from exploration missions and from Martian meteorites for model calibration in the following paragraphs.

We compared the elemental and isotopic compositions obtained from our model to those of the present-day atmosphere on Mars (Figures 6-11). Elemental abundances are compared in Figures 6a, 6b, and 6c. Because the atmospheric C ($CO_2$) abundance in the collapsed cases is determined by the equilibrium with the surface reservoirs in our model, the abundance always equals the observed value. The atmospheric C on present-day Mars was mainly sourced from the volcanic degassing. The baseline model (no comets) contained lower amounts of N and noble gases than the observed values (Figure 6a). In addition, lighter elements (Ne and Ar) were depleted compared with the chondritic or volcanic elemental pattern as a result of the sputtering and photochemical escape. Changing the volcanic degassing rate $C_{vol}$ mainly affected the N abundance (Figure 6a). Increasing $C_{vol}$ to 5 resulted in an adequate amount of N, whereas noble gases were still depleted compared with the present-day Mars. This is because the volcanic gas is depleted in noble gases (Figure 2). Hereafter we assume $C_{vol}$ = 5.

Next, we changed $C_{Ne,IDP}$ and $f_{comet}$ to reproduce the noble gas abundances assuming their exogenous origins. Because Ne was rapidly lost from a thin atmosphere due to the sputtering, Ne supplied by impacts has been lost in the present-day Mars. The lifetime of Ne in the Martian atmosphere was estimated to be ~0.1 Gyrs (Jakosky et al. 1994). The present-day abundance is controlled by continuous supply by IDPs. The model assuming Ne-rich IDPs is more suitable for reproducing the abundance in the Martian atmosphere (Figure 6b). We found that IDPs are only related to the Ne abundance (and vice versa) and do not affect the other elements in our model. We note that the origin of atmospheric Ne—derived from IDPs in our model—may depend on the assumption for the history and composition of volcanic degassing.

Apart from Ne, the observed noble gas abundances can be explained by a small amount of cometary contribution (Figure 6b). The suitable ratio of cometary to asteroid impacts ($f_{comet}$) was ~0.1%. The Ne abundance was insensitive to $f_{comet}$ because the episodically supplied Ne by cometary impacts was rapidly removed from the Martian atmosphere. Hereafter, we refer to the model assuming $C_{vol} = 5$, $C_{Ne,IDP} = 10$, and $f_{comet} = 0.1\%$, as the best-fit model (Figure 6c).

The best-fit model reproduced all of the observed isotopic compositions of the present-day atmosphere on Mars except for Xe isotopes. The evolution of the N isotopic composition is shown in Figure 7. The results of Monte Carlo simulations showed two different behaviors depending on whether the atmosphere collapsed or not (Figure 7a). Whereas the N isotope ratio preserved the primitive value under a dense atmosphere, the value started to increase as the atmosphere became thinner. The $^{15}N/^{14}N$ isotope ratio evolved stochastically in a collapsed thin atmosphere due to episodic impact events and the following escape-induced fractionation (Figure 7b). We note that the fractionation before 4.1 Ga was caused by the photochemical escape only (without sputtering) because the magnetic protection was assumed. After the quasi-steady state between ~3.5 Ga to ~0.5 Ga, the N isotope ratio started to increase as the volcanic degassing diminished. The N isotope ratio in the collapsed cases successfully reproduced the present-day value obtained by *Curiosity* (Wong et al. 2013).

The isotopic compositions of Ne ($^{20}Ne/^{22}Ne$), Ar ($^{38}Ar/^{36}Ar$), and Kr ($^{86}Kr/^{84}Kr$ as a representative ratio) in the Martian atmosphere were also reproduced in our model, while Xe ($^{136}Xe/^{130}Xe$ as a representative ratio) was not (Figures 8-11). As well as N, the evolutionary tracks of Ne and Ar are classified into two branches depending on the presence/absence of the atmospheric collapse (Figures 8a and 9a). In contrast to the isotope ratios moderately evolving in a dense atmosphere, stochastic fractionation was observed after the atmospheric collapse and the cessation of the dynamo assumed at 4.1 Ga (when the sputtering started to operate) because of episodic impact events and the following escape-induced fractionation. The averaged value of the present-day $^{38}Ar/^{36}Ar$ ratios in our simulations matched the data measured by *Curiosity* (Atreya et al. 2013). Our model also reproduced the $^{20}Ne/^{22}Ne$ ratio of the present-day atmosphere estimated from Martian meteorites (Pepin 1991).

Isotopic compositions of Kr and Xe in our model were determined by the mixing of the supplied values because of the inefficient sputtering due to their masses being heavier than N, Ne and Ar. Impacts eroded the original (chondritic) Kr and Xe isotopes and supplied the cometary (solar-like) isotopes in our best-fit model. We note that the isotopic composition at the end was chondritic in the cases where $f_{comet}$ is lower. The best-fit model is consistent with the solar-like isotopic pattern of Kr in the Martian atmosphere estimated from Martian meteorites (Pepin 1991) and recently measured by *Curiosity* (Conrad et al. 2016). We started our simulations by assuming that the isotopic compositions of Xe were identical to those in the present-day Martian atmosphere because the Xe isotopes have been considered as a remnant of early hydrodynamic loss (Pepin 1994). However, the initial isotopic composition of Xe was lost in the first $10^0$-$10^1$ Myrs in our model by the impact erosion and replenishment. Fractionation caused by Xe ionization on a timescale of billions years, which is poorly understood and is not included in our model, might be responsible for the fractionated Xe in the Martian atmosphere, as was proposed to explain fractionated Xe in the Earth's atmosphere (Pujol et al. 2011). We note that impactors having originally-fractionated Xe isotopes (Zahnle et al. 1990) cannot be ruled out.

The stochastic evolution of the isotopic compositions after the atmospheric collapse (Figures 7-11) was caused by stochastic impacts. The fluxes of the escape and replenishment processes are compared in Figure 12. The atmospheric erosion and replenishment by impacts were dominant processes prior to ~4 Ga. The supply of volatiles by the impacts led to the rise in both their abundances and escape rates in a collapsed atmosphere (Figures 12a and 12b), causing the following escape-induced fractionation (Figures 7b and 9b).

From the end of the LHB period (~3.5 Ga), the volcanic degassing dominated the replenishment of N and balanced the loss caused by the sputtering and photochemical escape (Figure 12a). This balance corresponds to the quasi-steady state observed in the N isotope ratio (Figure 7b). We note that the isotope ratio in the steady state can be written as (Yung et al. 1988; Jakosky et al. 1994; Kurokawa et al. 2016),

$$I_{m, n} = \frac{I_{m, n}^{0}}{R_{m/n}} . \qquad (12)$$

The increase of the $^{15}N/^{14}N$ ratio after ~0.5 Ga (Figure 7b) was caused by the decay of volcanic degassing and the dominance of the photochemical escape over the sputtering (Figure 12a). The photochemical escape induced more efficient fractionation than the sputtering (Section 2).

Stochastic impacts greatly affected the evolution of the Ar isotopes relative to N (Figures 12a and 12b); the major source of Ar is comets, in contrast to the N source being continuous volcanic degassing. This caused the $^{38}Ar/^{36}Ar$ ratio to be more scattered and a steady state to be unclear (Figure 9b).

In summary, the Martian atmosphere in our model evolved as follows. Impact erosion and replenishment governed the evolution before ~4 Ga, whereas the other processes (photochemical escape, sputtering, volcanic degassing and supply by IDPs) became effective after the LHB period. Changing $X_{gas}$ affected the balance between the impact erosion and replenishment during the early period. Major components of the atmosphere (C and N) on present-day Mars were mainly sourced from the volcanic degassing (controlled by $C_{vol}$). On the other hand, noble gases were supplied from comets and IDPs (controlled by $f_{comet}$ and $C_{Ne,IDP}$). Finally, the major finding of our model is the difference in the isotopic fractionation before and after the atmospheric collapse; a dense atmosphere preserved the primitive isotopic compositions, whereas a thin atmosphere experienced stochastic fractionation because of episodic supply of volatiles by impacts and the following escape-induced fractionation.

## 3.2. The effect of the atmospheric pressure on the timescales of the isotopic fractionation

The atmospheric pressure (namely, the total mass of the atmosphere) has a dominant effect on the degree of isotope fractionation at a given time. We show the influence of the atmospheric pressure on the timescales of the isotopic fractionation in this subsection. These are useful to understand our results and consequently to obtain constraints on the early atmosphere (subsection 3.3). We use N and Ar as examples because the *Curiosity* measurements are available and these isotopes will be used to constrain the atmospheric pressure on early Mars (subsection 3.3).

To demonstrate the effect of the atmospheric pressure on the fractionation timescales, we calculated the cases where the $CO_2$ partial pressure is forced to decrease gradually in the best-fit model (Figure 13). We do not specify the cause of the decrease in $CO_2$ pressure; both carbonate formation and $CO_2$ sublimation can contribute to the decrease. In these cases, we did not assume atmospheric collapse. The N ($^{15}N/^{14}N$) and Ar ($^{38}Ar/^{36}Ar$) isotope ratios increased faster as the atmosphere became thinner. Consequently, the stochastic behavior caused by impacts appeared. Primitive values were preserved over a longer timescale under a denser atmosphere. The onset of fractionation following the decrease in atmospheric pressure (Figure 13) and the rapid fractionation after the atmospheric collapse (Figures 7-11) can be explained by shorter timescales of the isotopic fractionation under a lower atmospheric pressure. The timescales of the fractionation induced by the photochemical escape and sputtering (the total number of the species in the atmosphere divided by the flux) are shown as functions of the atmospheric pressure and time in Figure 14. The timescales are shorter for a lower pressure and for an earlier time because of the smaller atmospheric volume and the higher EUV flux with more intense solar wind, respectively. Contrary to the timescales being $\sim 10^{-1}$-$10^{0}$ Gyrs for a dense atmosphere ($\sim$ several bar) at $\sim$4 Ga, those for a collapsed thin atmosphere (6 mbar in our model) are distinctly short ($\sim 10^{-3}$-$10^{-1}$ Gyrs). The decrease of the timescales caused the faster evolution in a thinner atmosphere (Figure 13) and distinct behaviors of the collapsed and uncollapsed atmospheres (Figures 7-11).

### 3.3. Constraining the atmospheric pressure on early Mars

We will constrain the atmospheric pressure on early Mars using the model calibrated by the elemental and isotopic compositions of the present-day atmosphere (the best-fit model, subsection 3.1). The best-fit model is further constrained by N and Ar isotopic compositions of an early atmosphere. We employed $\delta^{15}N$ and $^{38}Ar/^{36}Ar$ values of 7 ‰ and ≤0.2, respectively, for the 4 Ga atmospheric component (Mathew and Marti 2001). These values were obtained by stepwise-heating experiments of ALH 84001. ALH 84001 is a Martian orthopyroxinite that was initially formed by 4.1 Ga igneous activity (Bouvier et al. 2009; Lapen et al. 2010). ALH 84001 subsequently experienced a ~4 Ga secondary alteration process that formed carbonate (Borg et al. 1999); these carbonates are believed to have formed from water that was closely associated with the Noachian atmosphere (e.g., Niles et al., 2013, Halevy et al. 2011). Based on detailed investigations of trapped gas components in the individual steps, Mathew and Marti (2001) concluded that the $\delta^{15}N$ and $^{38}Ar/^{36}Ar$ values of 7 ‰ and ≤0.2 should represent an ancient (~4 Ga) atmospheric component.

The comparison of our model to the ALH 84001 data suggests an uncollapsed, moderately dense atmosphere (> 0.5 bar) at 4.1 Ga regardless of the presence/absence of the magnetic field at that time. The N and Ar isotope ratios at 4.1 Ga were compared to the simulated isotope ratios in various evolutionary tracks obtained by the best-fit model in Figures 7b, 9b, and 13. The $^{38}Ar/^{36}Ar$ isotope ratio started to increase because of the sputtering after both the atmospheric collapse and cessation of the magnetic dynamo (Figure 9b). This suggests that the atmosphere was not collapsed (> 0.5 bar, Forget et al. 2013) and moderately dense (> $\sim10^0$ bar, considering the timescale shown in Figure 14c); otherwise the sputtering was prohibited prior to 4.1 Ga. On the contrary, the $\delta^{15}N$ value started to increase, even before the cessation of the dynamo at 4.1 Ga, as the atmospheric pressure decreased (Figure 7b). The photochemical escape removed N and fractionated the N isotope ratio regardless of the presence/absence of the magnetic dynamo. The unfractionated N isotope ratio recorded in ALH 84001 was reproduced in the cases where the atmosphere was not collapsed (> 0.5 bar, Figure 7b) and moderately dense (> $\sim10^{-1}$-$10^0$ bar, see Figure 13 and Figure 14c). We note that the actual atmospheric pressure at 4.1 Ga should have been greater than this lower limit (0.5 bar) when several factors to increase the isotope ratios—for instance, the sputtering prior to 4.1 Ga—are considered (see subsection 4.3 and Figure 15).

Our results on the N and Ar isotope compositions suggested the lowest bounds (~0.5 bar) for the atmospheric pressure at 4.1 Ga. We emphasize that this lower limit was derived both from two independent constraints. One was that the timescales of escape-induced fractionation should be long enough to preserve the primitive isotopic compositions (Figures 13 and 14). The other was that the atmospheric pressure should be higher than the threshold of the atmospheric collapse (Figures 7 and 9). Therefore, the lower limit is valid even if the threshold of the atmospheric collapse is much lower or absent because of high obliquity, for instance (Forget et al. 2013; Soto et al. 2015).

# 4. Discussion

## 4.1. The evolution of the atmospheric pressure through time

A moderately dense atmosphere at ~4 Ga has also been suggested by other geochemical, geological, and theoretical constraints (Manga et al. 2012; Van Berk et al. 2012; Cessata et al. 2012; Kite et al. 2014; Forget et al. 2013; Hu et al. 2015). Constraints on the atmospheric pressure on early Mars are summarized in Figure 16. The lower limit obtained from this study (0.5 bar at 4.1 Ga) is consistent with most of these previous works.

Whereas our model predicted that the Martian atmosphere was denser than ~0.5 bar at 4.1 Ga, the atmospheric and surface reservoirs on present-day Mars contain $CO_2$ that corresponds to less than 100 mbar (Kurahashi-Nakamura and Tajika 2006; Lammer et al. 2008) (Figure 16). It suggests a subsequent decrease in the surficial $CO_2$ mass after 4.1 Ga. As the surficial $CO_2$ becomes less than ~0.5 bar, the atmosphere would have collapsed at once. Subsequent reinflation events might have occurred driven by obliquity changes (Forget et al. 2013; Kite et al. 2014). The inflated intervals, combined with other short-lived greenhouse effects (Segura et al. 2002; 2008), might be related to surface liquid water production in the Hesperian (Kite et al. 2014).

Comparisons to studies about the evolution of water on Mars (e.g., Kurokawa et al. 2014, 2016; Mahaffy et al. 2015; Villanueva et al. 2015) provide comprehensive views on the evolution of the surface environment on Mars. The evolution of the hydrogen isotope (D/H) ratio has suggested significant loss prior to 4.1 Ga (Kurokawa et al. 2014), probably because of thermal escape induced by strong EUV irradiation from the young sun. The EUV irradiation would also cause the photochemical escape of N as was assumed in this study. The decrease in the atmospheric pressure and the following atmospheric collapse would cause a climate change on Mars. The climate change might lead to a limited water exchange between the surface and subsurface reservoirs, resulting in heterogeneous D/H distribution on Mars (Mahaffy et al. 2015; Usui et al. 2015; Kurokawa et al. 2016).

*4.2. Comparisons with previous studies on the atmospheric pressure on early Mars*

The lower limit of ~0.5 bar on the atmospheric pressure at the time of ALH 84001 crystallization might be inconsistent with the upper limit of 0.4 bar that was argued to have existed at the same time by Cessata et al. (2012) (Figure 16). They used the high $^{40}Ar/^{36}Ar$ ratio (626 ± 100) of the trapped gas in ALH 84001 (Cessata et al. 2010) to evaluate the early atmospheric pressure and obtained this upper limit. Contrary to Cessata et al. (2012), in which a simplified evolution model was used, Slipski and Jakosky (2016) recently studied the evolution of Ar isotopic compositions ($^{36}Ar$, $^{38}Ar$, and $^{40}Ar$) with a more complicated model by modeling various processes and reservoirs. The high $^{40}Ar/^{36}Ar$ ratio at 4.1 Ga obtained by Cessata et al. (2010) was not reproduced in Slipski and Jakosky (2016). Slipski and Jakosky (2016) discussed several possibilities to reproduce the high $^{40}Ar/^{36}Ar$ ratio, including early catastrophic degassing and sputtering prior to 4.1 Ga, but neither Cessata et al. (2012) nor Slipski and Jakosky (2016) investigated these. These possibilities may be capable of reproducing the high $^{40}Ar/^{36}Ar$ ratio in a moderately dense atmosphere. In addition, we note that Mathew and Marti (2001) reported a much lower $^{40}Ar/^{36}Ar$ ratio (~130 ± 6) in the temperature step corresponding to the highest $^{38}Ar/^{36}Ar$ ratio. To identify the actual $^{40}Ar/^{36}Ar$ ratio at the time of ALH 84001 crystallization is crucial for constraining the early evolution of the Martian atmosphere.

Our model showed that a dense $CO_2$ atmosphere prevents minor atmospheric species from escaping, which is consistent with the behavior of the model of Slipski and Jakosky (2016). However, we found stochastic evolution of the $^{38}Ar/^{36}Ar$ ratio in contrast to the rather monotonic evolution in Slipski and Jakosky (2016). This is mainly caused by a difference in Ar sources: stochastic cometary impacts in our model and volcanic degassing in Slipski and Jakosky (2016). Cometary supply was not included in Slipski and Jakosky (2016). Because the volcanic gas is depleted in noble gases—even with $C_{vol} = 5$, which can supply enough N to the atmosphere in our model—we conclude that noble gases are likely to have exogenous origins (subsection 3.1). The exogenous origin of noble gases has also been proposed by other studies (e.g., Marty et al. 2016). We note that the monotonic evolution of the $^{38}Ar/^{36}Ar$ ratio can also be obtained in our model (Figure 13).

We assumed that the hydrodynamic escape did not occur after the transition to the $CO_2$-rich atmosphere. In contrast, Tian et al. (2009) proposed a hydrodynamic escape of the $CO_2$-rich atmosphere and concluded that a $CO_2$-dominated atmosphere could not have been maintained before 4.1 Ga. The contrast is due to the different exobase temperature; we adopted the exobase temperature of 1200 K at 4.5 Ga, which was calculated by assuming a $CO_2$-rich atmosphere and a solar XUV flux 100 times higher than the present-day value (Kulikov et al. 2007; Terada et al. 2009), while Tian et al. (2009) obtained >2500 K for a solar EUV flux 20 times larger than the present-day value. Adopting our model in the scenario of Tian et al. (2009) would lead to significant enrichment of heavy isotopes similar to the collapsed cases in this study (Section 3). Future measurements of Martian meteorites will provide atmospheric isotope data for different eras (see subsection 4.4) and enable us to test the various scenarios.

### 4.3. Validity of model assumptions

There are several $CO_2$ sinks, which are not included in our model, and parameters that are not investigated in detail. Carbonate formation (Hu et al. 2015; Wray et al. 2015) and basal melting of the $CO_2$ ice cap (Kurahashi-Nakamura and Tajika 2006; Longhi 2006) may have affected the evolution of the atmospheric pressure. The size and velocity distributions of impactors, the assumed temperature, and the total mass of impactors also affect the evolution. Changing these values would change the best-fit parameters moderately. However, our conclusion is not influenced by these changes, as the atmospheric pressure determines the time scale of the isotopic fractionation (subsection 3.2). Therefore, the lower limit of early atmosphere obtained in this study ( ~0.5 bar) is robust as long as the unfractionated N and Ar in ALH 84001 (Mathew and Marti 2001) actually represent the early atmosphere.

The actual atmospheric pressure at 4.1 Ga should have been greater than our lower limit (~0.5 bar) when we consider several factors to increase the isotope ratios. We show the evolution of N and Ar isotope ratios in the cases where the magnetic field was absent from the beginning (i.e., the sputtering operated before 4.1 Ga) in Figure 15. The sputtering fractionated these isotope ratios before the collapse even occurred. It suggests that a much denser atmosphere is required to explain the unfractionated isotope data at 4.1 Ga if the ancient magnetic field was weak or absent. We assumed that asteroids have a relatively low $\delta$ N value (-30‰, Table 4), but asteroids with higher δN values might have contributed to the late accretion. In addition, cometary impacts prior to 4.1 Ga are not included in our model. If these impactors with heavier N isotopic composition contributed to the late accretion, they would add to the increase in the N isotope ratio in the Martian atmosphere.

We calculated the evolution of the Martian atmosphere from 4.5 Ga by assuming that the hydrogen-rich primordial atmosphere has been lost by hydrodynamic escape at 4.5 Ga (Section 2). We test the validity of this assumption. The age of the solar system estimated from calcium-aluminum-rich inclusions is 4.567 Ga (Amelin et al. 2010; Connelly et al. 2012). The accretion timescale of Mars was estimated to be $1.8^{+0.9}_{-1.0}$ Myrs (Dauphas and Pourmand 2011). The mass of the primordial atmosphere on Mars is < $10^{18}$ kg (Ikoma and Genda 2006). Mars started to lose its primordial atmosphere after the dissipation of the solar nebula. The timescale of the dissipation of the solar nebula is a few Myrs (Haisch et al. 2001). The timescale to remove the hydrogen-rich atmosphere can be estimated by combining the analytical estimate of the escape rate and the estimated EUV flux at the time (e.g., Kurokawa and Kaltenegger 2013; Kurokawa and

Nakamoto 2014). The energy-limited formula of the hydrodynamic escape (e.g., Watson et al. 1981) is given by,

$$\frac{dM_{\text{atm}}}{dt} = \epsilon \frac{\pi F_{\text{EUV}} R_{\text{p}}^3}{G M_{\text{p}}}, \qquad (12)$$

where $M_{\text{atm}}$ is the atmospheric mass, $\epsilon$ is the efficiency ($\sim$0.1), $F_{\text{EUV}}$ is the solar EUV flux, $R_{\text{p}}$ and $M_{\text{p}}$ are the radius and mass of Mars, and $G$ is the gravitational constant. Using the solar EUV flux in the saturation phase estimated from solar proxies (Ribas et al. 2005, Section 2), the escape timescale is estimated to be $\sim 10^0$ Myrs. Therefore, it is valid to assume that the hydrogen-rich primordial atmosphere has been lost at 4.50 Ga or even earlier. We note that starting from an earlier time (e.g., from 4.56 Ga) requires higher atmospheric pressure at 4.1 Ga because the earlier EUV irradiation could affect N isotopes. Therefore, our constraint shown in Section 3 (> 0.5 bar) should be treated as a lower limit.

### 4.4. Synergy to the MAVEN mission and future analyses of Martian meteorites

Our model is capable of connecting the atmospheric escape modeling with geochemical records (isotopic data). The atmospheric escape rates from early Mars remain uncertain; the estimates on the sputtering rate vary in the previous studies (e.g., Luhmann et al. 1992; Chassefière et al. 2007; Chassefière and Leblanc 2011). The ongoing *MAVEN* mission (Jakosky et al. 2015) will provide better understanding of escape mechanisms and estimates on the escape rates from early Mars by observing the response of the Martian atmosphere to solar activity. Incorporating the escape rates into our model will provide further constraints on the evolution of the Martian atmosphere.

The isotopic compositions of N and noble gases recorded in ALH 84001 are crucial to constrain the atmospheric pressure on early Mars. Whereas we used the less-fractionated $^{15}\text{N}/^{14}\text{N}$ ratio reported by Mathew and Marti (2001), the data on N isotopic composition in ALH 84001 vary in the literature (up to > 150‰, Grady et al. 1998; Miura and Sugiura 2000). Mathew and Marti (2001) interpreted these heavy N components as cosmogenic. Identification of the more reliable N isotopic ratio at 4.1 Ga would help constrain the early evolution of the Martian atmosphere.

Since the reliability of our model depends largely on the 4 Ga atmospheric component, the model will be further tested by new measurements of Martian meteorites that will provide atmospheric isotope data for different eras. Because the crystallization age of ALH 84001 corresponds to the time of the LHB, we cannot rule out the possibility that the N and Ar signatures are sourced by a major impact, although any impacts observed in our model did not completely reset the isotopic compositions in the atmosphere. For example, newly found Northwest Africa (NWA) 7533 and NWA 7034 contain volatile-rich Martian regolith clasts with ages of 4.4 Ga and 2.1 Ga (Hymayun et al. 2013; Agee et al. 2013). Moreover, recent *in-situ* isotope measurements by ion microprobe indicate that alteration veins in 1.3 Ga nakhlites have hydrogen isotopic compositions similar to that of Hesperian-aged mudstone measured by *Curiosity* (Mahaffy et al. 2015), but distinct from that of the present-day atmosphere (Hallis et al. 2012). If possible, extraction of such ancient atmospheric components from meteorites by novel analytical approaches will offer a significant contribution toward the accurate understanding of the evolution of the Martian atmosphere.

# 5. Conclusions

The history of the loss and replenishment of the Martian atmosphere was simulated to constrain the atmospheric pressure on early Mars. The evolution of the isotopic compositions of N and noble gases was calculated by taking various processes into consideration: the atmospheric collapse, impacts of asteroids/comets, atmospheric escape induced by solar radiation/wind, volcanic degassing, and gas deposition by IDPs. Our model was capable of reproducing the elemental and isotopic compositions in the present-day atmosphere, except for Xe isotopic composition, for which an additional process that could be related to ionization, was suggested. In our best-fit model, atmospheric N originated mainly from volcanic degassing, whereas noble gases had exogenous origins: comets and IDPs. The calibrated model was adapted to early Mars to constrain the atmospheric pressure. We found that photochemical escape and sputtering efficiently fractionate the isotopes in a thin atmosphere during the early stage. We conclude that Mars had a moderately dense atmosphere (the atmospheric pressure is higher than approximately 0.5 bar) at 4 Ga, if the unfractionated N and Ar isotopes recorded in ALH 84001 represent the atmospheric components at the time. This lower limit is valid regardless of the presence/absence of the intrinsic magnetic field at the time because photochemical escape removed atmospheric N. Further constraints will be obtained by combining our model with upcoming *MAVEN* results and analyses of isotopic compositions in Martian meteorites formed in different ages.

# Acknowledgments

We thank B. M. Jakosky and an anonymous reviewer for comments and suggestions. We have greatly benefited from discussion in the workshop on planetary impacts held at the Institute of Low Temperature Science, Hokkaido University. HK is supported by JSPS KAKENHI Grant (15J09448). KK is supported by JSPS KAKENHI Grant (15H01067, 26610184, and 25871212). TU is supported by JSPS KAKENHI Grant (26800272, 15KK0153, and 16H04073).

**Figure 1:**

A schematic view of the model used in this study.

**Figure 2:**

Elemental abundances in asteroids, comets, IDPs, volcanic gas, and the Martian atmosphere, with respect to the solar values. The abundance in the atmosphere is normalized by the mass of solid Mars. Those of asteroids, comets, and IDPs ($C_{Ne,IDP}$ = 1) are normalized by the masses of these bodies. The abundance in volcanic gas shows the sum of degassed mass during 4.5 billion years in our model ($C_{vol}$ = 1) normalized by the mass of solid Mars. Data are from: solar abundance (Pepin 1991), asteroids (CI chondrites, Pepin 1991), comets (Dauphas 2003), IDPs (Flynn 1997), and the Martian atmosphere (Pepin 1991; Mahaffy et al. 2013).

**Figure 3:**

(a) Size and (b) velocity distributions of impactors (selected 20 examples of the Monte Carlo results).

**Figure 4:**

Volcanic degassing rates of $CO_2$ (purple), $N_2$ (green), $H_2O$ (sky blue) as a function of time. The model is based on Craddock and Greeley (2009) ($C_{vol}$ = 1).

**Figure 5:**

The evolution of the atmospheric pressure. (a) Volatile-rich impactor ($X_{gas}$ = 10%) cases and (b) volatile-poor impactor ($X_{gas}$ = 1%) cases. We assumed $C_{vol}$ = 1, $C_{Ne,IDP}$ = 1, and $f_{comet}$ = 0.1% in both cases (a) and (b). A hundred Monte Carlo simulations are shown. Collapsed and uncollapsed cases are shown by thick colored lines and thin black lines, respectively. (c) Selected three typical evolutionary tracks: the case where the atmosphere collapsed after 4.1 Ga (purple), the case where the atmosphere collapsed prior to 4.1 Ga (green), and the uncollapsed case (black). We assumed $X_{gas}$ = 1%, $C_{vol}$ = 5, $C_{Ne,IDP}$ = 10, and $f_{comet}$ = 0.1% for case (c).

**Figure 6:**

Simulated elemental abundances in our model with respect to solar values. Average values of the collapsed cases are shown. (a) Purple: $C_{vol}$ = 1, $C_{Ne,IDP}$ = 1, and $f_{comet}$ = 0%, green: $C_{vol}$ = 3, $C_{Ne,IDP}$ = 1, and $f_{comet}$ = 0%, sky blue: $C_{vol}$ = 5, $C_{Ne,IDP}$ = 1, and $f_{comet}$ = 0%, respectively. The inner subset shows an enlarged view of the nitrogen abundance. (b) Purple: $C_{vol}$ = 5, $C_{Ne,IDP}$ = 1, and $f_{comet}$ = 0%, green: $C_{vol}$ = 5, $C_{Ne,IDP}$ = 10, and $f_{comet}$ = 0%, sky blue: $C_{vol}$ = 5, $C_{Ne,IDP}$ = 10, and $f_{comet}$ = 0.1%, yellow: $C_{vol}$ = 5, $C_{Ne,IDP}$ = 10, and $f_{comet}$ = 1%, respectively. (c) Our best-fit model: $C_{vol}$ = 5, $C_{Ne,IDP}$ = 10, and $f_{comet}$ =

0.1%, with 1-σ error bars. We assumed $X_{gas}$ = 1% in all cases. The abundances in the Martian atmosphere are the same as Figure 2 (red).

**Figure 7:**
The evolution of the nitrogen isotope ratio (the $\delta^{15}N$ value) in the best-fit model ($X_{gas}$ = 1%, $C_{vol}$ = 5, $C_{Ne,IDP}$ = 10, and $f_{comet}$ = 0.1%). (a) A hundred Monte Carlo simulations. (b) Selected three cases corresponding to Figure 5c. Collapsed and uncollapsed cases are shown by thick colored lines and thin black lines, respectively. The averaged value of the collapsed cases at 0 Ga is shown with 1-σ error bar (purple). Ranges of observed present-day atmospheric ratios are from Curiosity (red solid-line, Wong et al. 2013), Viking (red dashed-line, Nier and McElroy 1977), and ALH 84001 (red data point at 4.1 Ga, Mathew and Marti 2001), respectively.

**Figure 8:**
The evolution of the Ne isotope ratio ($^{20}Ne/^{22}Ne$ value) in the best-fit model ($X_{gas}$ = 1%, $C_{vol}$ = 5, $C_{Ne,IDP}$ = 10, and $f_{comet}$ = 0.1%). (a) A hundred Monte Carlo simulations. (b) Selected three cases corresponding to Figure 5c. Collapsed and uncollapsed cases are shown by thick colored lines and thin black lines, respectively. The averaged value of the collapsed cases at 0 Ga is shown with 1-σ error bar (purple). A range of observed present-day atmospheric ratio (red) is from Martian meteorites (Pepin 1991).

**Figure 9:**
The evolution of the Ar isotope ratio ($^{38}Ar/^{36}Ar$ value) in the best-fit model ($X_{gas}$ = 1%, $C_{vol}$ = 5, $C_{Ne,IDP}$ = 10, and $f_{comet}$ = 0.1%). (a) A hundred Monte Carlo simulations. (b) Selected three cases corresponding to Figure 5c. Collapsed and uncollapsed cases are shown by thick colored lines and thin black lines, respectively. The averaged value of the collapsed cases at 0 Ga is shown with 1-σ error bar (purple). Ranges of observed present-day atmospheric ratios are from Curiosity (red solid-line, Atreya et al. 2013), Viking (red dashed-line, Owen et al. 1977), and ALH 84001 (red data point at 4.1 Ga, Mathew and Marti 2001), respectively.

**Figure 10:**
The evolution of the Kr isotope ratio ($^{86}Kr/^{84}Kr$ value as an example) in the best-fit model ($X_{gas}$ = 1%, $C_{vol}$ = 5, $C_{Ne,IDP}$ = 10, and $f_{comet}$ = 0.1%). (a) A hundred Monte Carlo simulations. (b) Selected three cases corresponding to Figure 5c. Collapsed and uncollapsed cases are shown by thick colored lines and thin black lines, respectively. The averaged value of the collapsed cases at 0 Ga is shown with 1-σ error bar (purple). A range of observed present-day atmospheric ratio (red) is from Martian meteorites (Pepin 1991).

**Figure 11:**
The evolution of the Xe isotope ratio ($^{136}Xe/^{130}Xe$ value as an example) in the best-fit model ($X_{gas}$ = 1%, $C_{vol}$ = 5, $C_{Ne,IDP}$ = 10, and $f_{comet}$ = 0.1%). (a) A hundred Monte Carlo simulations. (b) Selected three cases corresponding to Figure 5c. Collapsed and uncollapsed cases are shown by thick colored lines and thin black lines, respectively.

The averaged value of the collapsed cases at 0 Ga is shown with 1-σ error bar (purple). A range of observed atmospheric ratio (red) is from Martian meteorites (Pepin 1991).

**Figure 12:**
The escape or replenishment rate of each process for (a) N and (b) Ar. The case where the atmosphere collapsed after 4.1 Ga (presented by a purple line in Figure 5c) is shown. The values are positive and negative for processes indicated by plus and minus, respectively. As the erosion and replenishment by impacts are treated as instantaneous processes in our model, these rates shown in the figures are the number of supplied/removed particles divided by the time duration from one impact to the next.

**Figure 13:**
The cases where the $CO_2$ partial pressure was assumed to decrease exponentially with the timescale $\tau_{CO2}$. (a) The assumed evolution of $CO_2$ partial pressure and the evolution of (b, c) N and (d, e) Ar isotope ratio (the $\delta^{15}N$ and $^{38}Ar/^{36}Ar$ values). A hundred Monte Carlo simulations are averaged. We also plotted the 1-σ dispersion in (c,e) with dashed lines. Colors and thicknesses in (b-e) correspond to those of the pressure evolution in (a).

**Figure 14:**
The contours of the timescales of (a) the photochemical escape of N, (b) the sputtering of N, and (c) the sputtering of Ar, respectively, as functions of the time and atmospheric pressure. The timescale is defined as the total number of particles divided by the flux. We assumed that only two components are present: $CO_2$ and the species considered. The mixing ratio of the species to $CO_2$ is assumed to be, $N_{N2}/N_{CO2} = 10^{-2}$ and $N_{Ar}/N_{CO2} = 10^{-6}$, respectively. We note that the dependence on the mixing ratio is weak.

**Figure 15:**
The evolution of the (a) N and (b) Ar isotope ratio (the $\delta^{15}N$ and $^{38}Ar/^{36}Ar$ values) in the best-fit model ($X_{gas} = 1\%$, $C_{vol} = 5$, $C_{Ne,IDP} = 10$, and $f_{comet} = 0.1\%$) in the case where sputtering was assumed to start from 4.5 Ga.

**Figure 16:**
Constraints on the atmospheric pressure on Mars obtained from this study (red line) and other studies (black lines). References are (1) this study, (2) Manga et al. (2012), (3) van Berk et al. (2012), (4) Cessata et al. (2012), (5) Kite et al. (2014), (6) Hu et al. (2015), (7) Forget et al. (2013) and (8) Kurahashi-Nakamura and Tajika (2006). Possible range of the evolutionary track is suggested by a blue shaded area.

Table 1: Sputtering yields (particles ejected per ions incident) divided by the abundance at the exobase) $Y_i$

| Element | $Y_i$ |
|---------|-------|
| C ($CO_2$) | $0.7^a$ |
| N ($N_2$) | $2.4^a$ |
| Ne | $3^a$ |
| Ar | $1.4^a$ |
| Kr | $1.143^b$ |
| Xe | $0.738^b$ |

[a] Results of Monte Carlo calculations shown by Jakosky et al. (1994)

[b] Calculated from $Y_i$ of oxygen atoms by assuming that $Y_i \propto U_i^{-1}$ ($\propto m_i^{-1}$), where $U_i$ is the gravitational binding energy (Johnson 1992)

Table 2: The altitude difference between the homopause and exobase $\Delta z$ and the upper atmospheric temperature $T_{upper}$ under different EUV conditions

| EUV level [× present-day EUV flux at Mars orbit] | $\Delta z$ [km] | $T_{upper}$ [K] | $\Delta z / T_{upper}$ [km K$^{-1}$] |
|---|---|---|---|
| 100 | 600 [a] | 1200 [a] | 0.50 |
| 6 | 175 [b] | 500 [b] | 0.35 |
| 3 | 125 [b] | 300 [b] | 0.42 |
| 1 | 80 [c] | 200 [c] | 0.40 |

[a] Terada et al. (2009), [b] Zhang et al. (1993), [c] Jakosky et al. (1994)

Table 3: The rates of gas deposition by IDPs

| Element | Gas deposition rate [g Gyr$^{-1}$] |
|---|---|
| Ne | $4.89 \times 10^{12\,a}$ ($C_{\text{Ne,IDP}} = 1$), $4.89 \times 10^{13}$ ($C_{\text{Ne,IDP}} = 10$) |
| Ar | $1.11 \times 10^{12\,a}$ |
| Kr | $1.64 \times 10^{10\,a}$ |
| Xe | $7.33 \times 10^{9\,a}$ |

$a$: Flynn (1997)

Table 4: Isotopic compositions of the Martian atmosphere and those used in our model

| | Mars | Volcanic degassing | Asteroids | Comets | IDPs |
|---|---|---|---|---|---|
| $\delta^{15}N$ [‰] | $572 \pm 82$[a] | -30[b] | -30[c] | 1000[d] | - |
| $^{20}Ne/^{22}Ne$ | $10.1 \pm 0.7$[c] | 13.7[c] | 8.9[c] | 13.7[c] | 13.7[e] |
| $^{36}Ar/^{38}Ar$ | $4.2 \pm 0.1$[f] | 5.3[c] | 5.3[c] | 5.4[g] | 5.8[c] |
| $^{78}Kr/^{84}Kr$ | $(6.37 \pm 0.36) \times 10^{-3}$ | $5.962 \times 10^{-3}$ | $5.962 \times 10^{-3}$ | $6.470 \times 10^{-3}$ | $6.470 \times 10^{-3}$ |
| $^{80}Kr/^{84}Kr$ | $4.09 \times 10^{-2}$ | $3.919 \times 10^{-2}$ | $3.919 \times 10^{-2}$ | $4.124 \times 10^{-2}$ | $4.124 \times 10^{-2}$ |
| $^{82}Kr/^{84}Kr$ | $0.2054 \pm 0.0020$ | 0.20149 | 0.20149 | 0.20629 | 0.20629 |
| $^{83}Kr/^{84}Kr$ | $0.2034 \pm 0.0018$ | 0.20141 | 0.20141 | 0.20340 | 0.20340 |
| $^{86}Kr/^{84}Kr$ | $0.3006 \pm 0.0027$ | 0.30950 | 0.30950 | 0.29915 | 0.29915 |
| $^{124}Xe/^{130}Xe$ | $(2.45 \pm 0.24) \times 10^{-2}$ | $2.851 \times 10^{-2}$ | $2.851 \times 10^{-2}$ | $2.947 \times 10^{-2}$ | $2.947 \times 10^{-2}$ |
| $^{126}Xe/^{130}Xe$ | $(2.12 \pm 0.23) \times 10^{-2}$ | $2.512 \times 10^{-2}$ | $2.512 \times 10^{-2}$ | $2.541 \times 10^{-2}$ | $2.541 \times 10^{-2}$ |
| $^{128}Xe/^{130}Xe$ | $0.4767 \pm 0.0103$ | 0.5073 | 0.5073 | 0.50873 | 0.50873 |
| $^{129}Xe/^{130}Xe$ | $16.400 \pm 0.080$ | 6.358 | 6.358 | 6.287 | 6.287 |
| $^{131}Xe/^{130}Xe$ | $5.147 \pm 0.0037$ | 5.043 | 5.043 | 4.9958 | 4.9958 |
| $^{132}Xe/^{130}Xe$ | $6.460 \pm 0.088$ | 6.150 | 6.150 | 6.0479 | 6.0479 |
| $^{134}Xe/^{130}Xe$ | $2.587 \pm 0.024$ | 2.359 | 2.359 | 2.1288 | 2.1288 |
| $^{136}Xe/^{130}Xe$ | $2.294 \pm 0.024$ | 1.988 | 1.988 | 1.6634 | 1.6634 |

*a*: Wong et al. (2013), *b*: Mathew and Marti (2001), *c*: Pepin (1991), *d*: Füri and Marty (2015), *e*: Pepin et al. (2000), *f*: Atreya et al. (2013), *g*: Balsiger et al. (2015)

Krypton and xenon data are from Pepin (1991) and references therein.

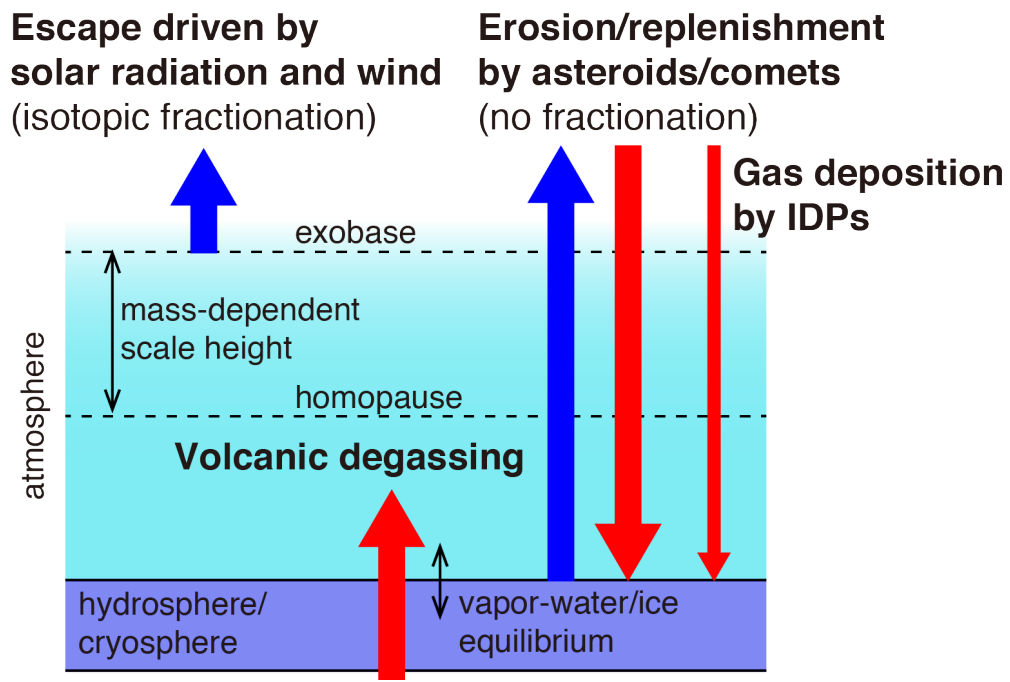

Figure 1

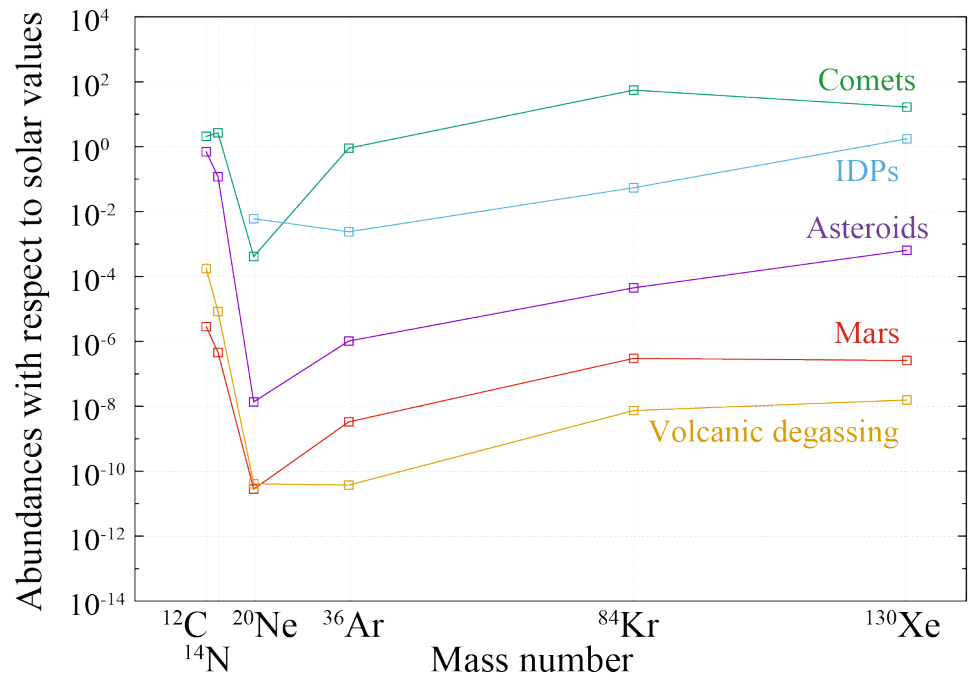

Figure 2

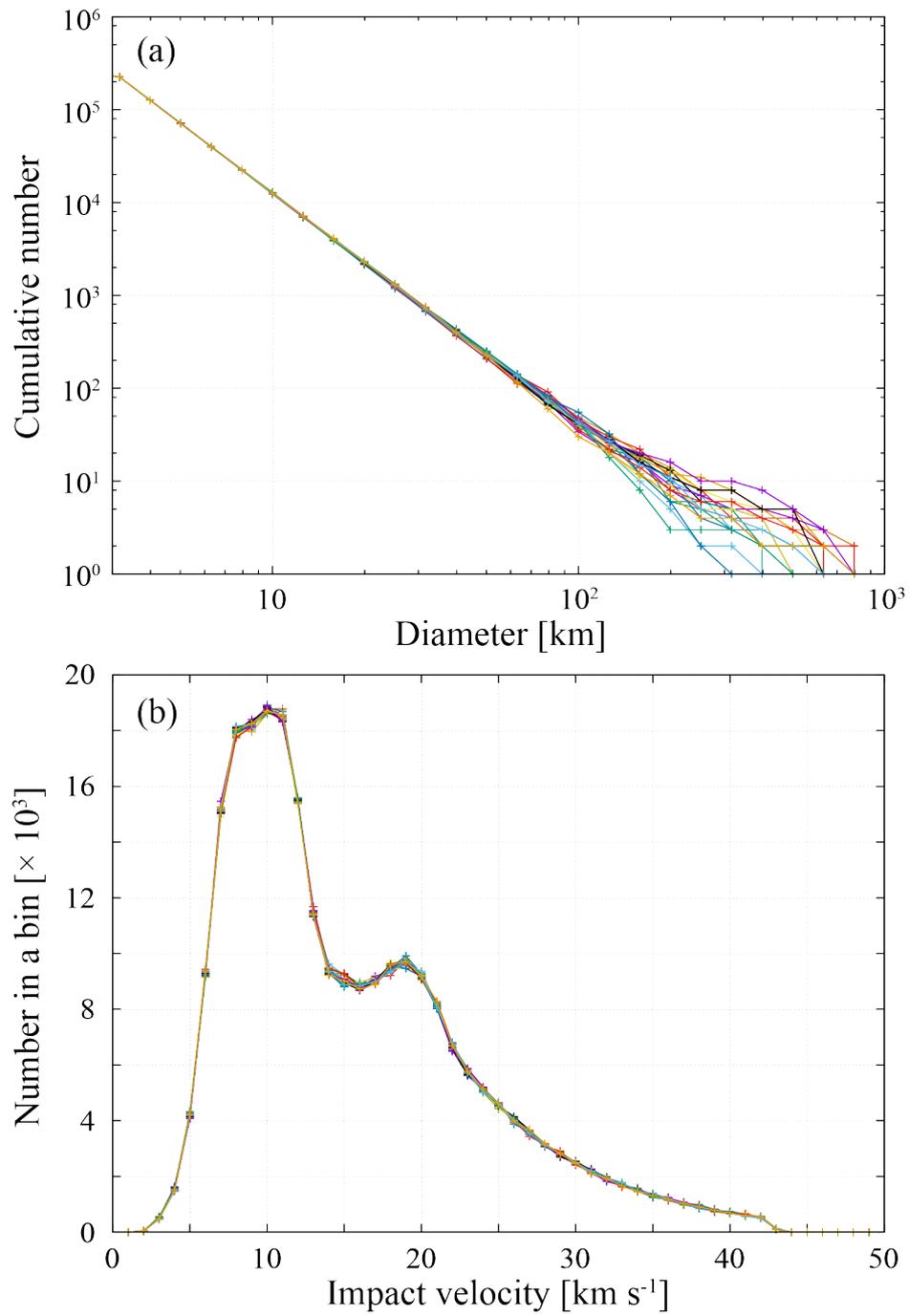

Figure 3

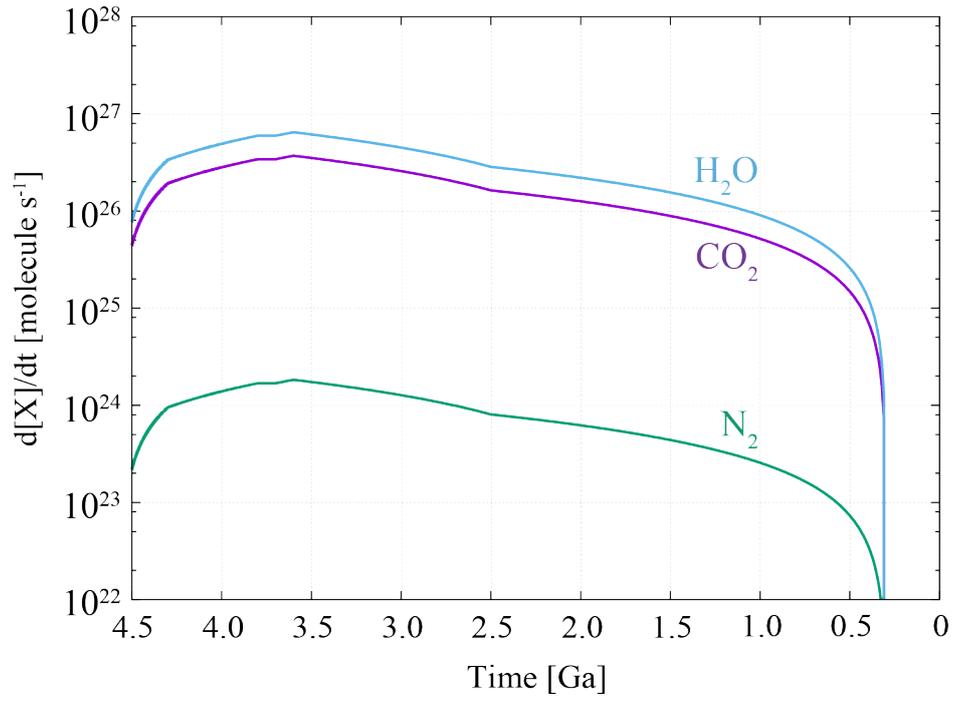

Figure 4

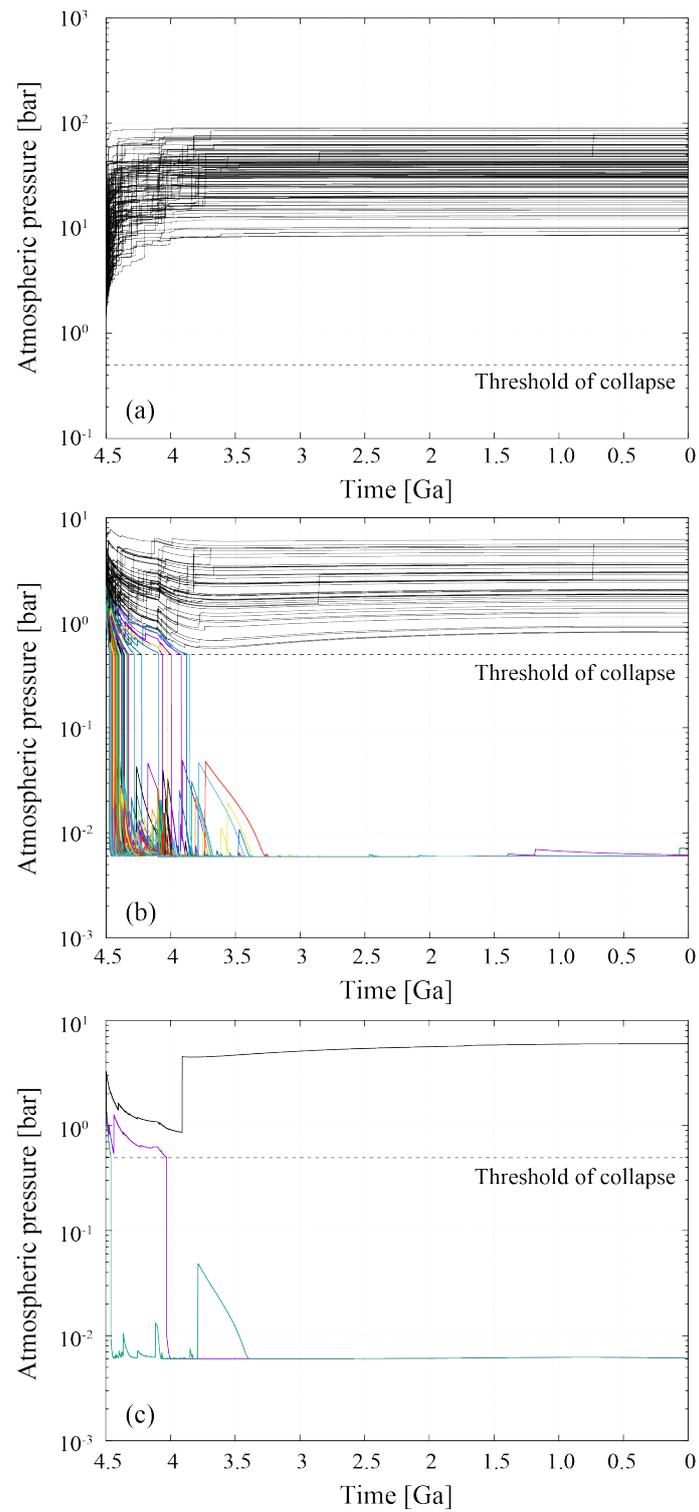

Figure 5

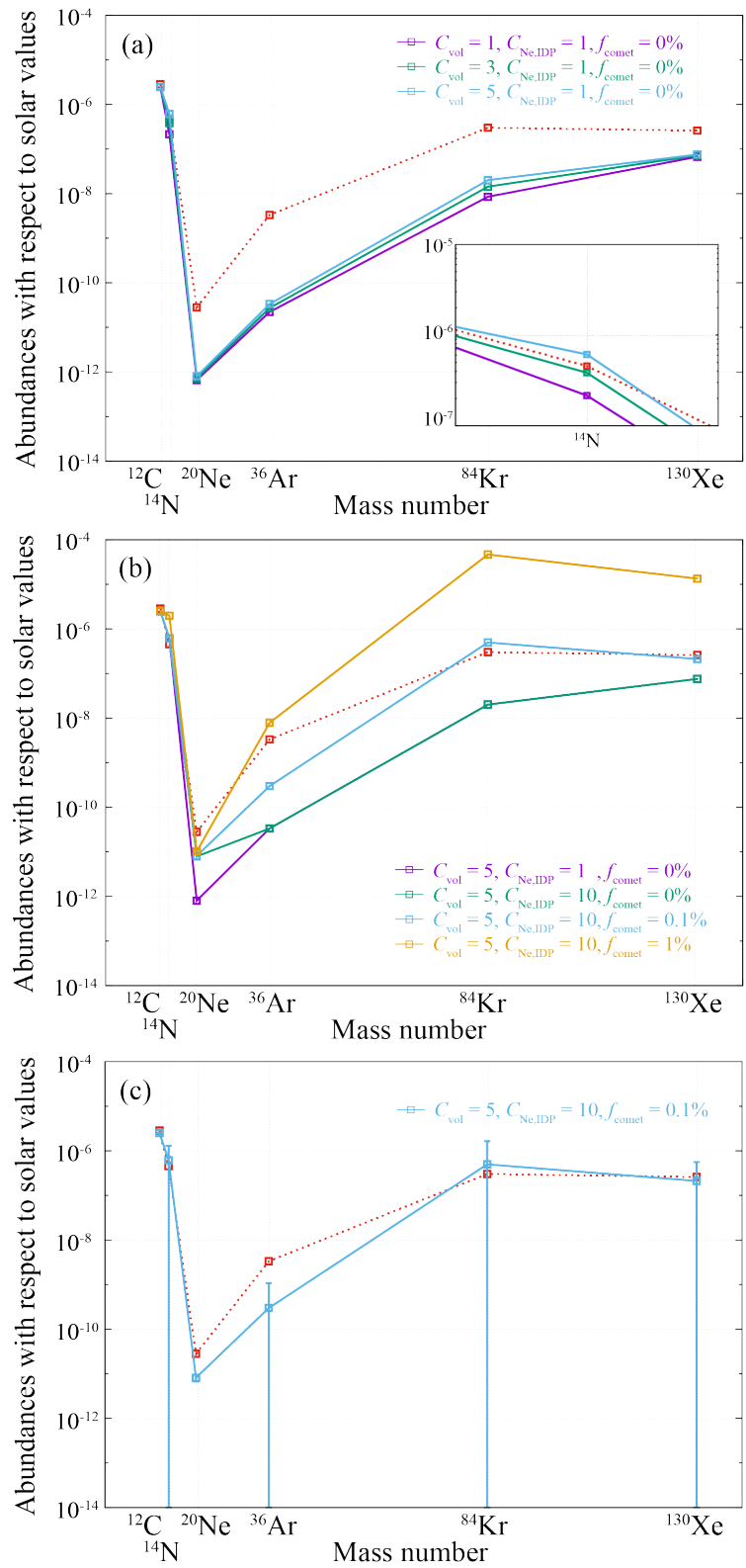

Figure 6

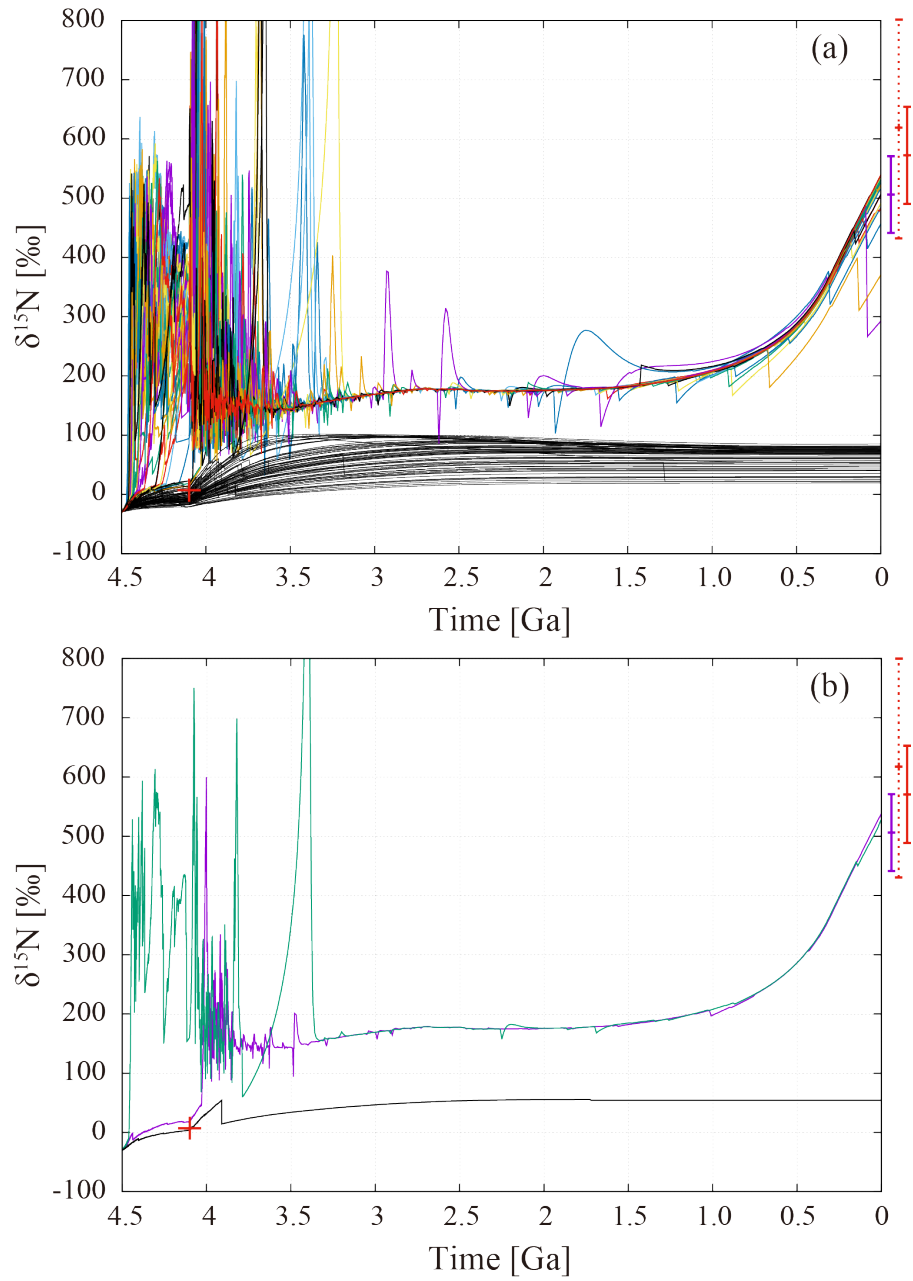

Figure 7

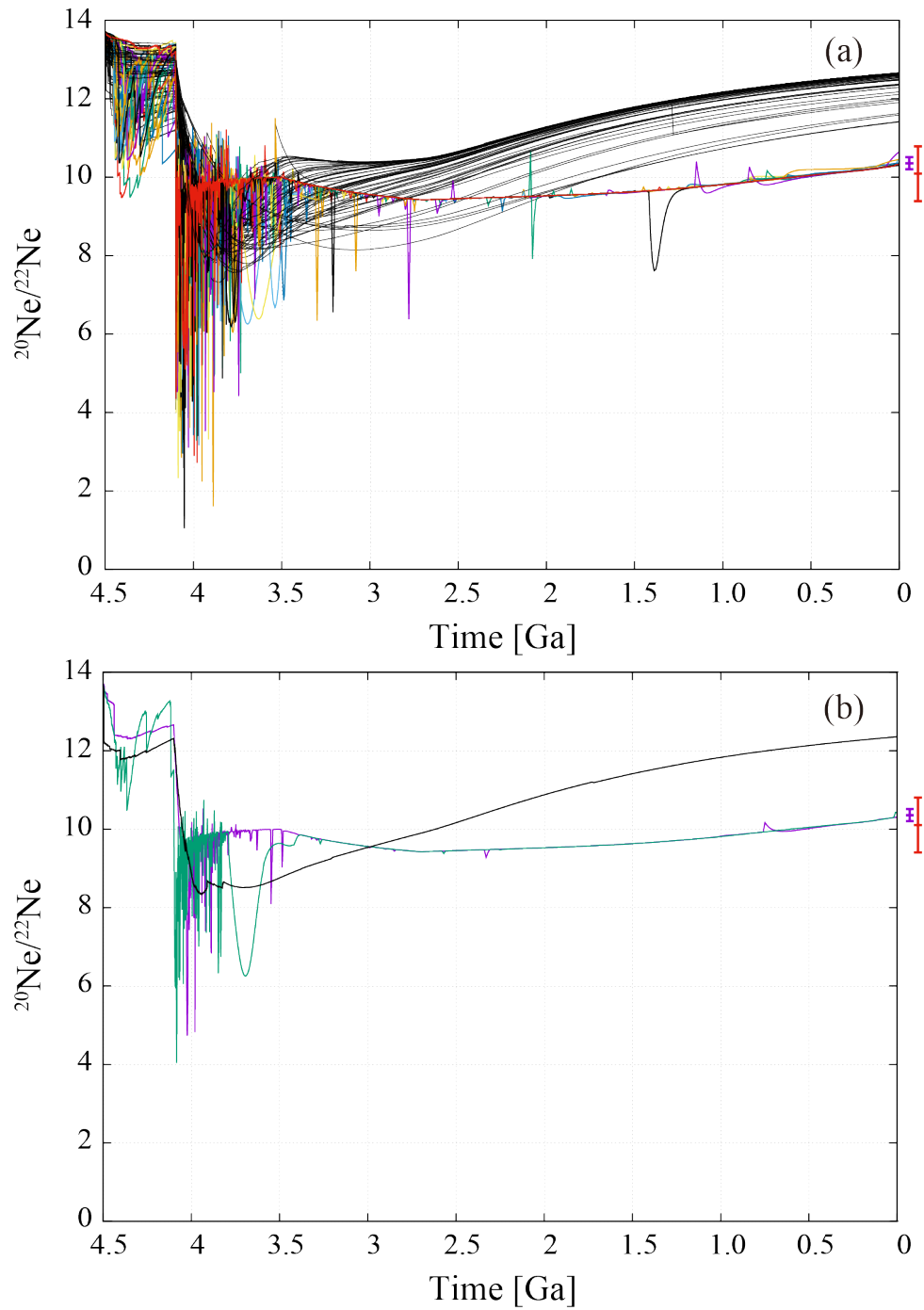

Figure 8

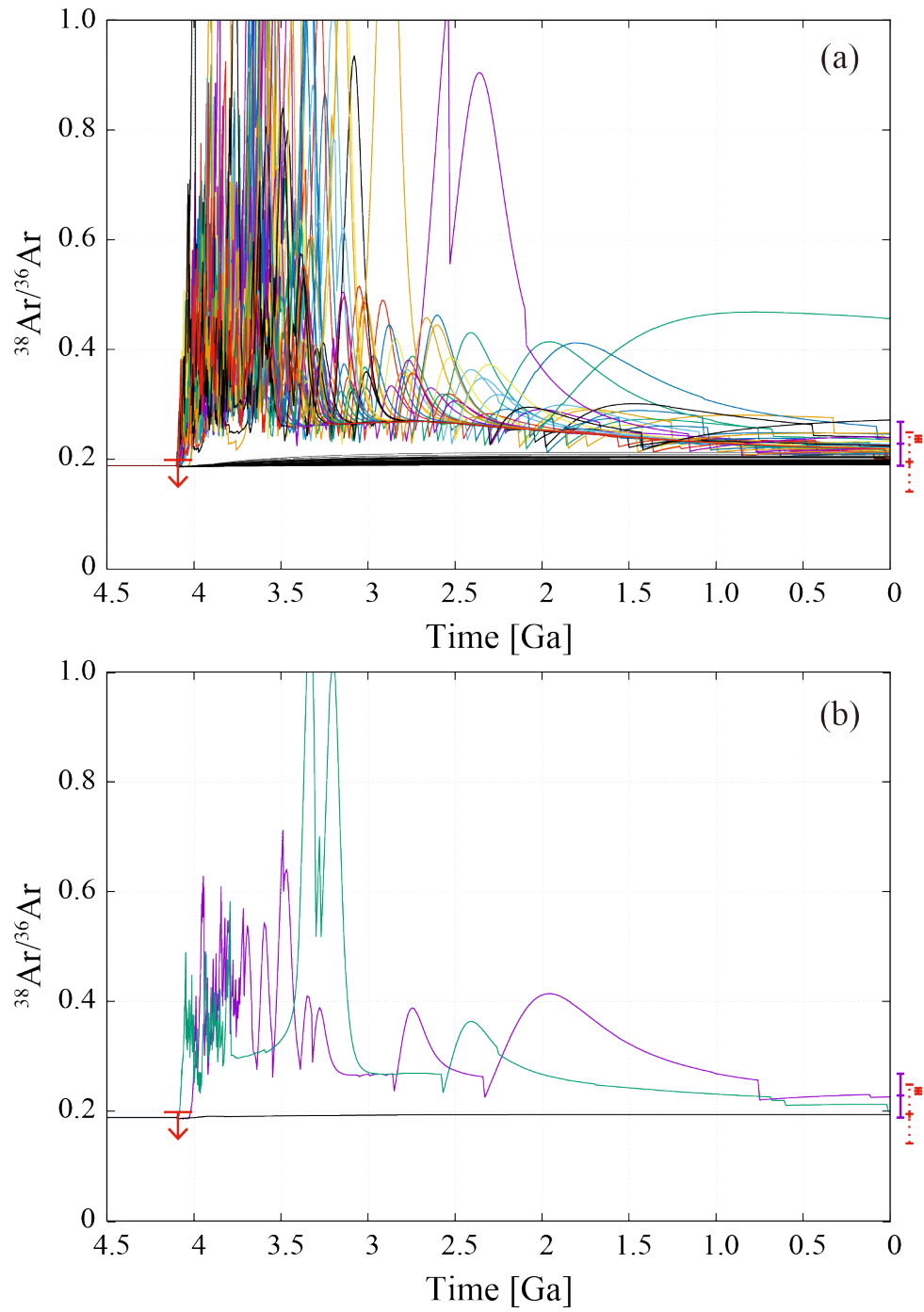

Figure 9

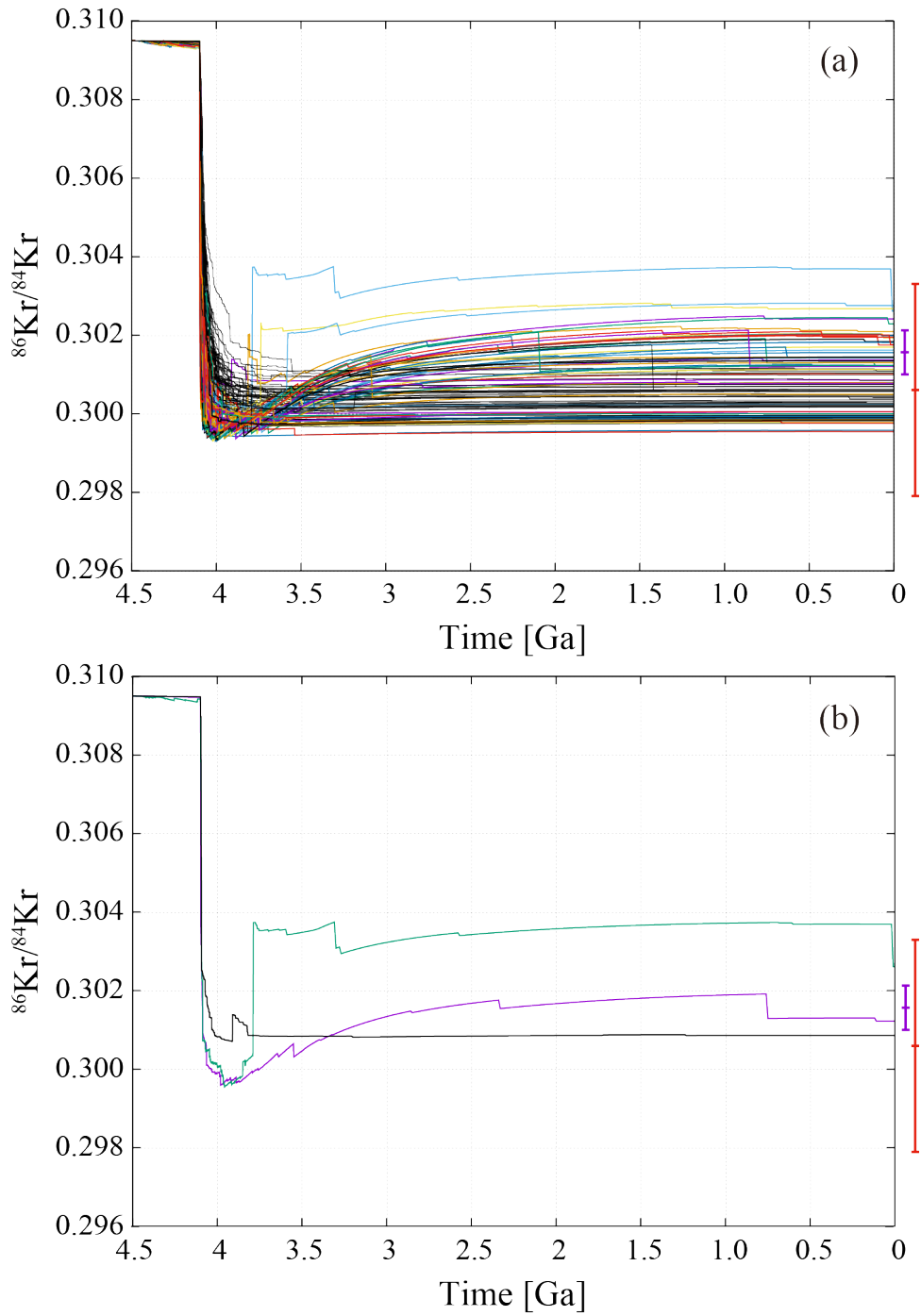

Figure 10

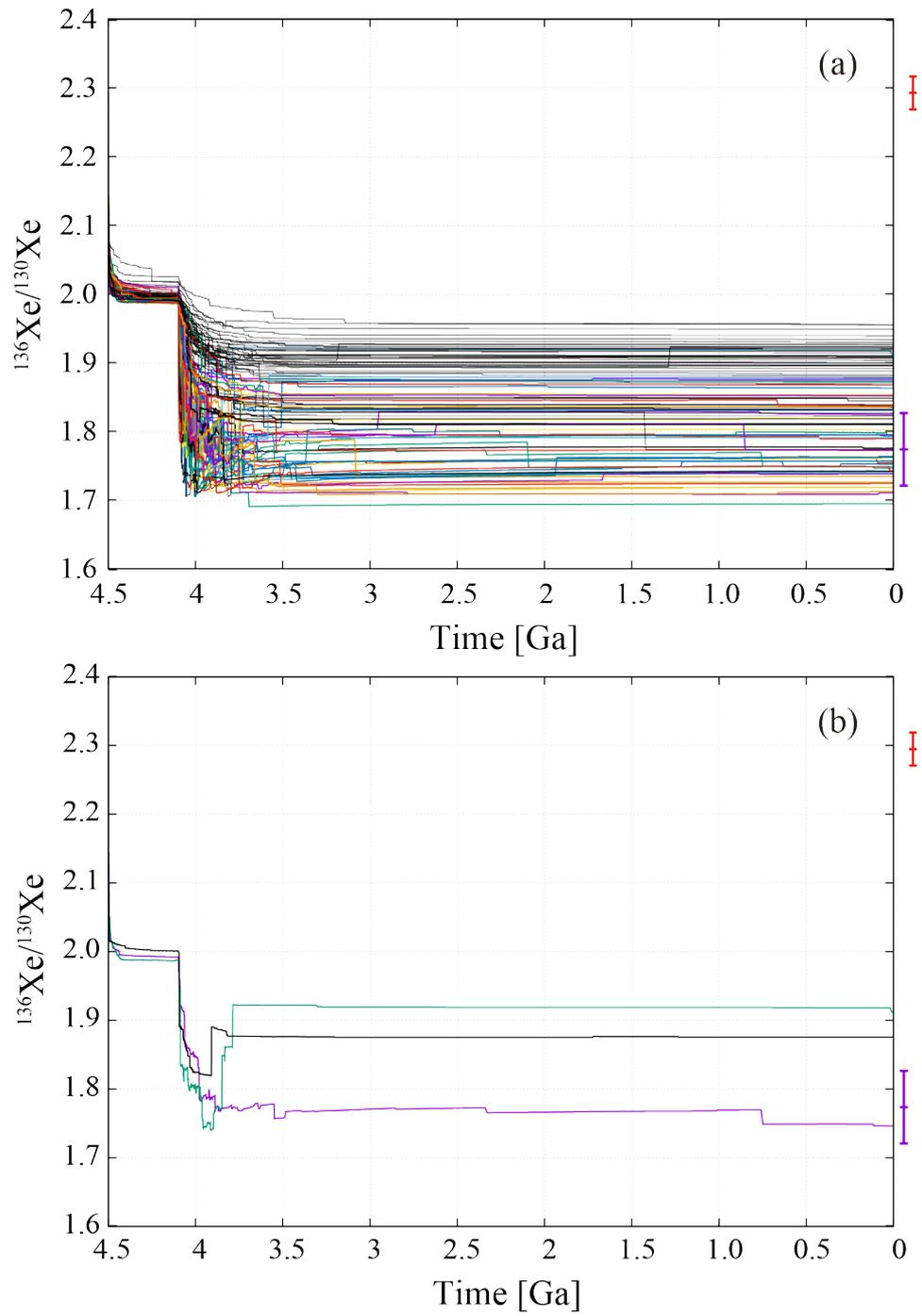

Figure 11

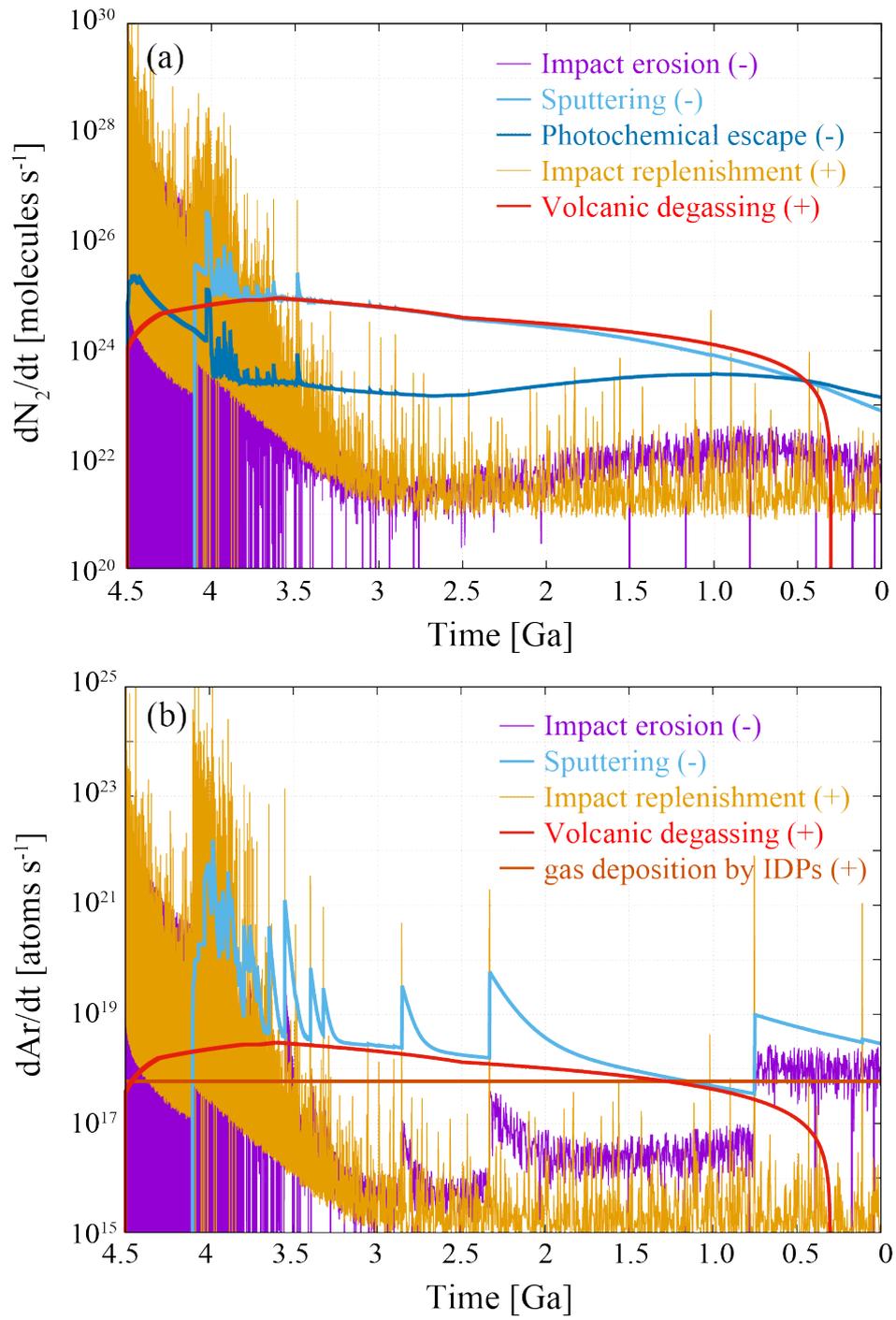

Figure 12

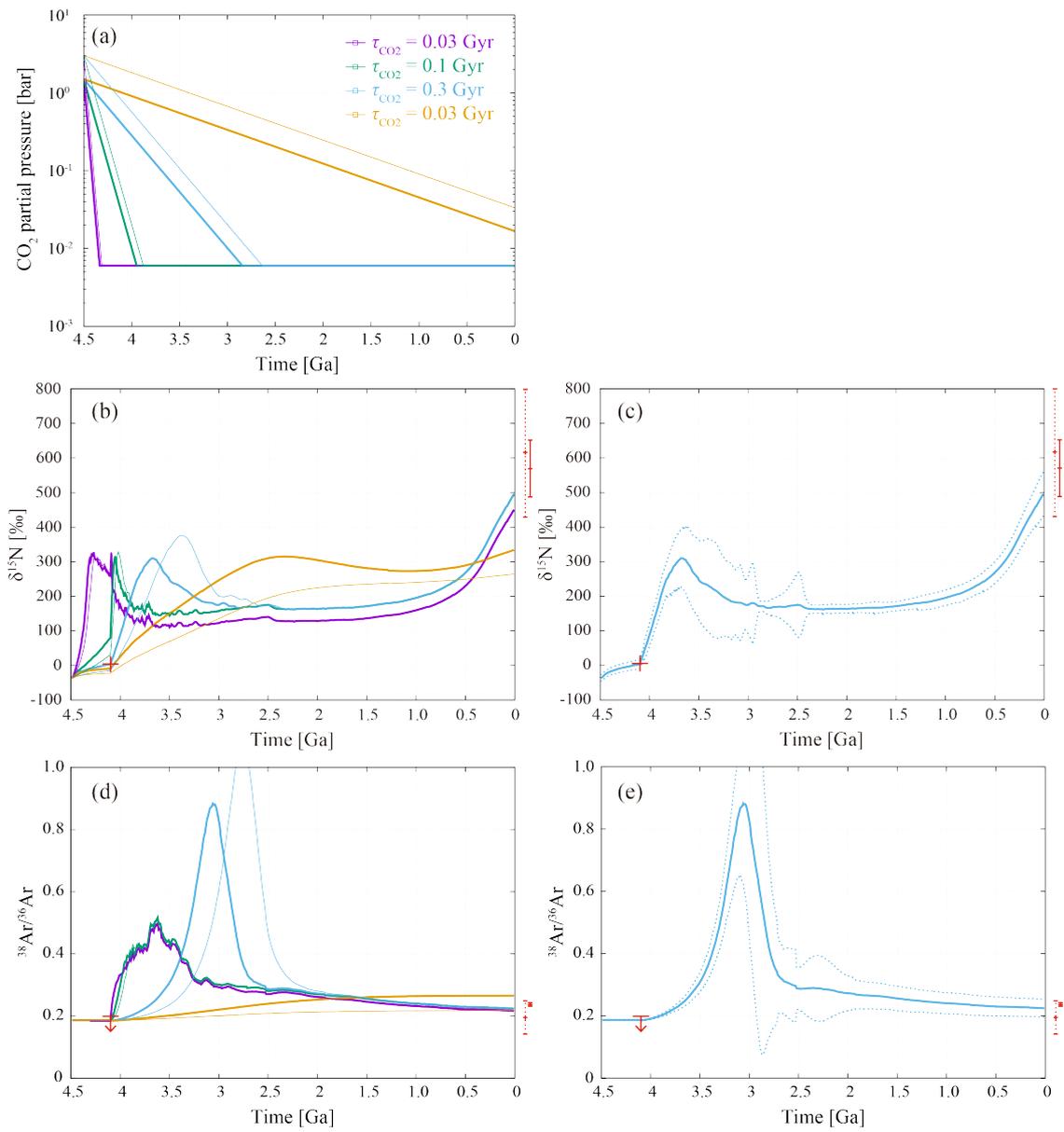

Figure 13

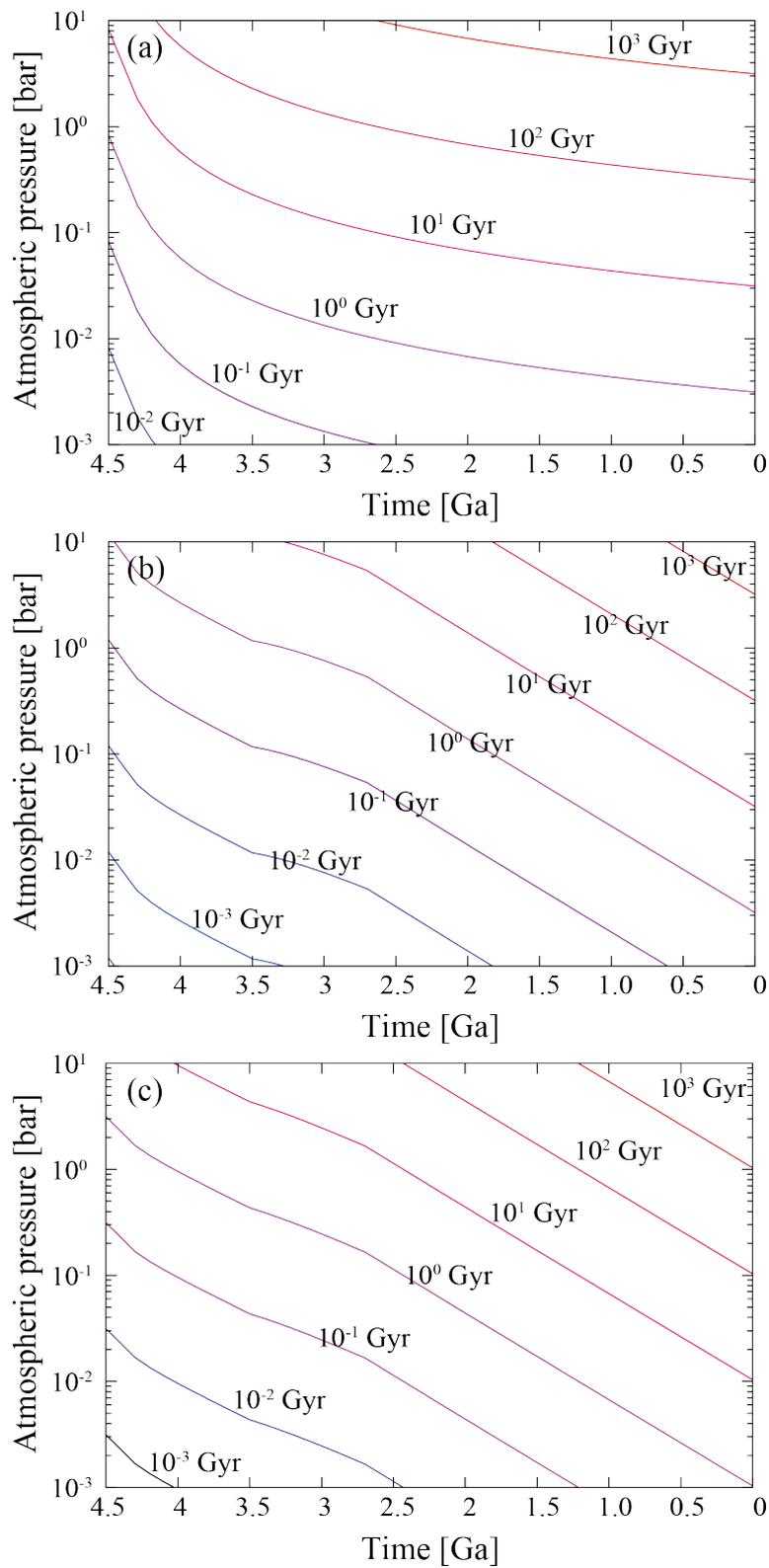

Figure 14

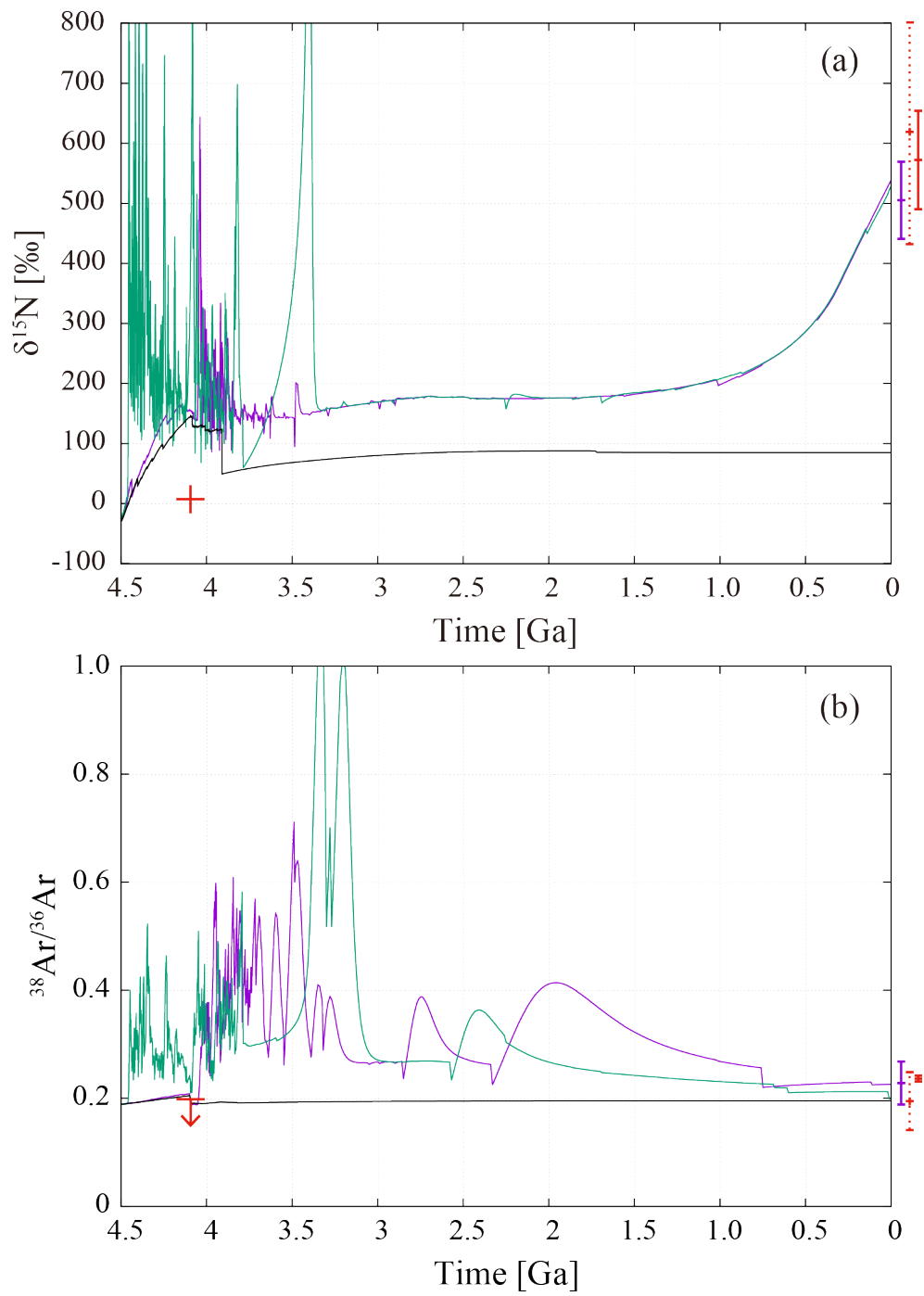

Figure 15

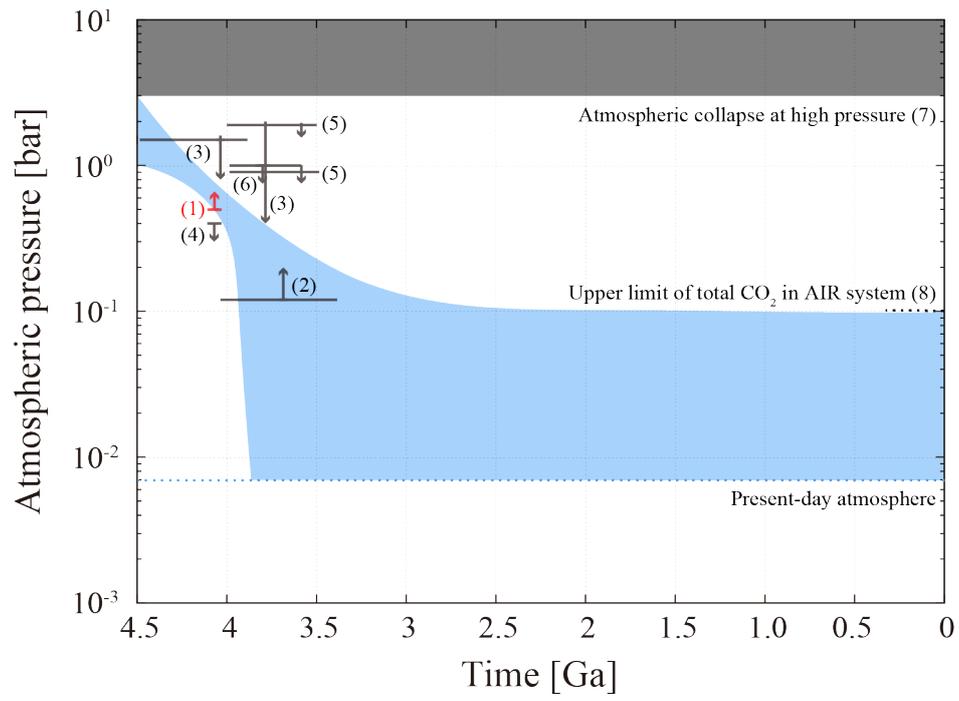

Figure 16